\documentclass[]{llncs} 
\usepackage{amsmath,amssymb,stmaryrd,mathtools,nccmath}

\usepackage{stmaryrd,mathtools} 

\usepackage{color,listings}

\usepackage{tikz}
\usetikzlibrary{positioning}  

\usepackage{booktabs}
\usepackage{multirow}
\usepackage{relsize}

\usepackage{microtype}
\usepackage[T1]{fontenc}
\usepackage{anyfontsize}

\usepackage{hyperref}  
\usepackage{cleveref}

\lstdefinelanguage{Maude}{%
   keywords={
    , mod, fmod, endm, endfm
    , pr , protecting 
    , ex , extending 
    , inc, including
    , sort, sorts, subsort, subsorts
    , var, vars
    , op, ops
    , eq, ceq
    , mb, cmb
    , rl, crl
    , class, subclass
    , if, then, else, fi
    , search
    , smt-search
    , such 
    , that
    },
    alsodigit={-},
    morekeywords = [2]{
     , ctor
     , assoc
     , id:
     , owise
     , comm
     , right
     , frozen}
}

\lstdefinelanguage{PLC}{%
   keywords={
    , PROGRAM, VAR, END_VAR, VAR_GLOBAL
    , RESOURCE, END_RESOURCE
    , DINT, REAL, BOOL, TIME
    , BEGIN, END_PROGRAM
    , FUNCTION_BLOCK , VAR_INPUT , VAR_INOUT, END_FUNCTION_BLOCK
    , END_VAR , VAR_OUTPUT, VAR_IN_OUT
    , IF, THEN , END_IF, ELSIF
    , CASE, OF, END_CASE
    , ELSE, WHILE, DO, END_WHILE
    , AND, NOT, OR
    , CONFIGURATION, END_CONFIGURATION,
    , TASK, INTERVAL, PRIORITY, WITH, ON
    }
}

\lstnewenvironment{maude}[1][\footnotesize]
{\lstset{language=Maude ,
  , basicstyle=\ttfamily#1 
  , commentstyle={}
  , columns=fullflexible
  , numbers=none
  , showstringspaces=false
  , keepspaces=true
  , escapechar=&
  , mathescape
  , frame = single
  , framexrightmargin=-3pt
  } 
}
{}

\lstnewenvironment{plc}[1][\footnotesize]
{\lstset{language=PLC ,
, basicstyle=\linespread{0.8}\ttfamily#1
, columns=fullflexible
, frame = single
, framexrightmargin=-3pt
} 
}
{}

\newcommand{\power}[1]{2^{#1}}
\newcommand{\enabled}[1]{\mathit{enabled}(#1)}
\newcommand{\ample}[1]{\mathit{ample}(#1)}
\newcommand{\AP}{\mathit{AP}}

\newcommand{\karrow}{\curvearrowright}
\newcommand{\kdots}{\mathop{...}}

\newcommand{\kframe}[1]{\framebox[1.0\linewidth]{\ensuremath{\begin{aligned} #1 \end{aligned}}}}

\newenvironment{krule}[2]
{
    \scriptstyle\texttt{#1:}
    \begin{array}{#2}
}
{
    \end{array}  
}

\usepackage[commandnameprefix=always]{changes} 
\usepackage{multirow}

\begin{document}
\graphicspath{{image/}}

\title{Formal Analysis of Networked PLC Controllers 
Interacting with Physical Environments}
\author{}
\institute{}
 \author{
 Jaeseo Lee\inst{1}  
 \and
 Kyungmin Bae\inst{1} 
 } 
 \institute{Pohang University of Science and Technology, 
 Korea}
\authorrunning{J. Lee et al.}

\maketitle


\begin{abstract}
Programmable Logic Controllers (PLCs) are widely used in industrial automation to control physical systems. 
As PLC applications become increasingly complex, ensuring their correctness is crucial. 
Existing formal verification techniques focus on individual PLC programs in isolation, often neglecting interactions with 
physical environments and network communication between controllers. 
This limitation poses significant challenges in analyzing real-world industrial systems, 
where continuous dynamics and communication delays play a critical role.
In this paper, 
we present a unified formal framework that integrates discrete PLC semantics, networked communication, 
and continuous physical behaviors. 
To mitigate state explosion, we apply partial order reduction, 
significantly reducing the number of explored states while maintaining correctness.
Our framework 
enables precise analysis of PLC-driven 
systems with continuous dynamics and networked communication. 

\end{abstract}

\section{Introduction}
\label{sec:intro}

Industrial automation systems rely on Programmable Logic Controllers (PLCs) to execute control tasks across 
various domains, including robotic assembly lines, power distribution systems, and autonomous production machinery. 
%
The IEC 61131-3 
standard \cite{international1993programmable}  
defines a set of domain-specific  languages 
that support PLC programming.
A PLC program typically controls a physical plant with continuous behaviors. 
Ensuring the correctness of PLC programs is crucial, as they are often deployed in safety-critical industrial environments. 

The need for rigorous verification of PLC programs has attracted significant research interest from both academia and industry.
As a result, various formal techniques and tools have been developed 
for formally analyzing PLC programs.
Representative works 
 include 
 \cite{darvas2015plcverif,gourcuff2006efficient,lamperiere2000formal,bauer2004verification,li2016automatic,rausch1998formal,bohlender2018cycle},
which address different aspects of PLC verification
and span various PLC programming languages.
Among these languages, Structured Text (ST),
a high-level imperative programming language,
is the most expressive and widely adopted for formal analysis 
\cite{darvas2017plc}.

However, existing work on formally analyzing PLC programs mostly focuses on a single PLC program in isolation, 
which neglects interactions with their physical environments and networked communication between multiple PLCs. 
Some approaches attempt to integrate physical dynamics and networked communication, but they often rely on abstracted 
PLC models that omit key programming constructs, such as function blocks and structured control flow. 
As a result, they fail to capture the full expressiveness of PLC languages, 
which are crucial for accurately  capturing real-world industrial automation systems.

The goal of this paper is to develop formal analysis techniques for general PLC applications 
by integrating three important aspects: 
the semantics of PLC programming languages, 
networked communication between PLC controllers with explicit consideration of network delays,
and
the behavior of physical plants with continuous dynamics.
Instead of addressing 
these aspects separately, we aim to develop a unified formal framework
that supports faithful reasoning about system-level properties involving all three aspects. 
%

Achieving this goal is challenging. Traditionally, each aspect of PLCs has been modeled and analyzed 
using a different formalism,
lacking an integrated approach for the full PLC-controlled system:

\begin{itemize}
    \item \textbf{PLC programs}: 
    Translated into the input language of another model checker \cite{gourcuff2006efficient,darvas2015plcverif,bohlender2018cycle},
    or given formal semantics of the language directly \cite{wang2023k,huang2019kst,plcst2022lee};
    
    \item \textbf{Communications}: 
    Modeled using communication models such as process calculi \cite{lanotte2020process},
     and colored Petri nets \cite{ghanaim2011modeling}; and

    \item \textbf{Physical Environments}: 
    Typically using hybrid automata \cite{hassapis1998validation,nellen2016}, 
    where PLC program behaviors are encoded as a guarded transition system.

\end{itemize}

Combining these approaches is nontrivial due to  high verification complexity from 
complex PLC programs, interleavings from asynchronous communication, 
and continuous 
behaviors.
For example, hybrid automata can model physical environments and
communications, but only under the unrealistic assumption 
that PLC programs---%
with features of general-purpose imperative languages---%
are simplified to finite state machines.
Even with such assumptions, 
the resulting models often become infeasible for formal analysis
(see \Cref{sec:expr}).

This paper
proposes a novel integration---tailored to the PLC domain---of three well-established techniques:
(i) object-oriented modeling of PLC programs, communication networks, and physical dynamics within a unified rewriting logic framework;
(ii) symbolic reachability analysis using Maude combined with SMT solving \cite{maude-se-2024} to handle continuous dynamics; and
(iii) partial order reduction to mitigate state-space explosion from concurrent interleavings.
While each technique has been studied individually,
the novelty of our approach lies in the nontrivial integration of all three within a single coherent framework.



We provide an \emph{object-oriented rewriting-based specification}, where each PLC controller, 
interacting with a physical plant and communicating with other PLC controllers, is encapsulated as an object,
based on rewriting logic \cite{MESEGUER199273}.
The attributes of a PLC object include timers, physical parameters, and a ``processor'' containing the entire configuration 
of a PLC ST program. The behavior of these objects is specified using an existing rewriting-based semantics for PLC ST \cite{wang2023k,huang2019kst}, 
supporting a full subset of the PLC ST language with minimal modification.


A communication model for PLCs with continuous behaviors can be effectively captured 
using \emph{distributed  real-time object-based systems} \cite{olveczky2007semantics}. 
This approach provides a structured way to represent interactions between PLC controllers and their physical 
environment. 
As is common in distributed systems, communication between objects is modeled as asynchronous message passing.
%
A key advantage 
is that it enables a direct and faithful representation of PLC communication mechanisms as specified in industrial standards.

To enable efficient verification, we apply partial order reduction (POR) \cite{Peled2018}
to mitigate the state explosion 
caused by redundant interleavings. 
Consider two PLC objects, 
each running a simple PLC program. Due to interleaving, the number of possible 
execution sequences grows exponentially. 
However, only statements that send and receive messages can produce different outcomes.
%
By integrating POR with symbolic rewriting-based model checking,
we significantly reduce the number of explored states while preserving correctness.

This integration is nontrivial because PLC systems exhibit unique features 
that are rarely addressed together by typical POR techniques. 
PLC programs 
use encapsulated function blocks with internal state,
execute in periodic scan cycles with strict real-time constraints,
and 
operate in distributed settings where each object interacts with physical plants.
We address these challenges by exploiting the modularity of rewriting-based specifications,
which facilitates the characterization of independent rewrite rules 
specialized for PLC communication models.

Our approach is implemented in Maude \cite{maude-book}, a tool for modeling and analyzing rewriting logic specifications.
We evaluate it on a set of  benchmark models, including \emph{new PLC ST benchmarks we developed} to explicitly incorporate 
PLC semantics, 
networked communication, and continuous dynamics---an integration not addressed in prior benchmarks. 
%
%
%
%
We compare our approach with a hybrid-automata-based tool, SpaceEx \cite{frehse2011spaceex}, showing that our method, combined with POR, 
outperforms SpaceEx by an order of magnitude. 

The key contribution of our paper
is to enable
\emph{unified formal analysis for PLCs} by tailoring the framework to the PLC domain,  
leveraging core concepts from the PLC standard, such as
sensing/actuation, scan cycles, and communication. 
%
%
While each underlying method---%
object-oriented modeling,
symbolic analysis,  and POR---%
has been explored in prior work,
their integration in the PLC context
has not been developed due to its inherent complexity.
%
We believe this work is the first to jointly address 
PLC scan-cycle semantics,
function block behavior, and networked physical interactions within a single formal analysis tool.

The rest of our paper is organized as follows. 
\Cref{sec:related} discusses related work. 
 \Cref{sec:prelim} provides background on 
 rewriting logic, PLC ST, and  POR. 
\Cref{sec:motiv} 
introduces a motivating example, involving PLC programs, physical environments,
and communication. 
\Cref{sec:external}
presents the object-oriented semantics
of PLCs with physical environments.
\Cref{sec:communication}
formalizes PLC communication.
%
\Cref{sec:reduction}
presents our POR method specialized for PLCs.
\Cref{sec:separation}
explains symbolic analysis using rewriting modulo SMT.
\Cref{sec:expr} reports the experimental results.
Finally, 
\Cref{sec:concl} presents some concluding remarks.



\section{Related Work}
\label{sec:related}

Various methods have been developed for the formal analysis of PLC programs written in different languages, including
Structured Text~\cite{gourcuff2006efficient,darvas2015plcverif,bohlender2018cycle,wang2023k,huang2019kst,plcst2022lee},
Function Block Diagram~\cite{li2016automatic,pavlovic2010model},
Ladder Diagram~\cite{rausch1998formal,lobov2004modelling},
Sequential Function Chart~\cite{lamperiere2000formal,hassapis1998validation,bauer2004verification},
and
Instruction List~\cite{canet2000towards}.
However, 
most of these approaches focus only on analyzing single PLC programs in isolation,
without considering interactions with physical environments or communication between multiple PLC controllers.
%
In contrast, our work explicitly models both physical environment interactions and PLC communication.


The papers \cite{hassapis1998validation,nellen2016} address the verification of PLC-controlled systems while incorporating physical dynamics.
These approaches employ a hybrid automata-based framework and introduce a tool capable of reasoning about both control programs and their surrounding physical environments.
In particular, \cite{nellen2016} utilizes a CEGAR approach, where the abstract model assumes arbitrary dynamics,
and refinement progressively adds concrete dynamics.
However, these works do not consider communication between different PLC controllers.
Additionally, they are restricted to SFCs,
whereas we consider ST (the most expressive of all PLC languages \cite{darvas2017plc}).
%

The work in \cite{lee2024fm} defines a formal semantics for PLC ST with preemptive multitasking using the K framework 
and introduces state space reduction techniques. 
However, it does not consider interactions with the physical environment or communication between PLC controllers. 
In contrast, our work focuses on multiple single-task PLC controllers that communicate with each other 
and interact with physical environments. 
While the paper \cite{lee2024fm} applies partial order reduction in the context of multitasking, 
we use it to address state explosion caused by communication. 
These differences suggest that our approach and the technique in \cite{lee2024fm} are complementary.

Only a few studies focus on the semantics of communicating PLCs, including \cite{lanotte2020process,ghanaim2011modeling}.
The paper \cite{lanotte2020process} defines a process calculus for runtime enforcement, enhancing security against malware,
and \cite{ghanaim2011modeling} models of PLC-based networked control systems using colored Petri nets.
A key similarity between their approaches and ours is the formal semantics for PLCs with communication.
However, 
both are based on highly abstracted formalisms,
lacking concrete semantics for any specific PLC programming language.
In contrast, our work explicitly considers both physical interactions and PLC communications in PLC ST.

Rewriting logic has been extensively applied to specifying and analyzing distributed object systems,
real-time and cyber-physical systems, and programming language semantics~\cite{meseguer2012twenty}.
%
%
In particular, rewriting modulo SMT~\cite{rw-smt-jlamp} has gained increasing attention as a symbolic analysis method.
Applications of rewriting modulo SMT include security protocols~\cite{8823756}, soft agents~\cite{nigam2022automating},
virtually synchronous cyber-physical systems~\cite{hybridsynchaadl,lee2022modeling}, autonomous robots~\cite{10.1007/978-3-319-66562-7_44},
business process models~\cite{duran2018symbolic}, and so on.
Within this research direction, our paper presents the first attempt to combine partial order reduction with symbolic reachability analysis.




\section{Preliminaries}
\label{sec:prelim}

\subsection{\bf Rewriting Logic and Maude}

Rewriting logic~\cite{unified-tcs} is a formal framework for specifying systems by representing system states as algebraic terms and describing their evolution through rewrite rules. A rewrite theory \cite{MESEGUER199273} $(\Sigma, E, R)$ consists of: 
(i) an equational theory $(\Sigma, E)$ specifying system states as algebraic data types, where $\Sigma$ is a signature (i.e., declaring sorts, subsorts, and function symbols) and $E$ is a set of equations; and
(ii) a set of rewrite rules $R$ of the form $l : t \rightarrow t'$ if $\mathit{condition}$, specifying the
system behavior, where $l$ is a label, and $t$ and $t'$ are terms. 
To specify real-time systems, we can add tick rewrite rules of the form
$\texttt{\{} t \texttt{\} => \{} t' \texttt{\} in time } \tau \texttt{ if } \mathit{cond}$ to model time elapse, where the
whole state has the form $\{t\}$ \cite{olveczky2006realtime}.


Rewriting modulo SMT describes a system as an evolution of \emph{constrained terms}. 
A constrained term
is a pair $\phi(x_1,\ldots,x_n) \parallel t(x_1,\ldots,x_n)$
of a constraint $\phi$
and a term $t$ over SMT variables $x_1, \ldots, x_n$ 
\cite{rw-smt-jlamp,bae2019symbolic}.
A constrained term $\phi \parallel t$ symbolically represents
the set of all instances of 
$t$ satisfying 
$\phi$,
denoted by $\llbracket \phi \parallel t \rrbracket$.
A \emph{symbolic rewrite}
$\phi_t \parallel t \rightsquigarrow_\mathcal{R}^\ast \phi_u \parallel u$
on constrained terms
symbolically represents
the set of
all ``concrete'' rewrites $t' \longrightarrow_\mathcal{R}^\ast u'$
such that
$t' \in \llbracket \phi_t \parallel t \rrbracket$ and $u' \in \llbracket \phi_u \parallel u \rrbracket$.
For any concrete rewrite $t' \longrightarrow_\mathcal{R}^\ast u'$
with $t' \in \llbracket \phi_t \parallel t \rrbracket$,
there exists
a symbolic rewrite 
$\phi_t \parallel t \rightsquigarrow_\mathcal{R}^\ast \phi_u \parallel u$
with $u' \in \llbracket \phi_u \parallel u \rrbracket$. Symbolic rewriting can be implemented using theory transformation as standard rewriting. For example, a rewrite rule of form $l : t \longrightarrow t'$ if $\psi$ can be transformed into $l : \texttt{PHI} \parallel t \longrightarrow (\texttt{PHI}\ \mathit{and}\ \psi \parallel t' \texttt{ if smtCheck(PHI}\ \mathit{and}\ \psi \texttt{)}$, where \texttt{PHI} is a Boolean expression and \texttt{smtCheck} is a function that checks the satisfiability of a given SMT formula.

Maude \cite{maude-manual} is a formal specification language based on rewriting logic, designed for both execution and analysis. A Maude module representing a rewrite theory $(\Sigma, E, R)$ specifies their conditional rewrite rules with the syntax $\texttt{crl [}l\texttt{] : } t \texttt{ => } t' \mbox{ if } \mathit{cond}$ (or, for unconditional rules, $\texttt{rl [}l\texttt{] : } t \texttt{ => } t')$, where $\mathit{cond}$ is a conjunction of equations. Similarly, equations are declared with the syntax $\texttt{ceq }t = t'\texttt{ if }\mathit{cond}$ (or $\texttt{eq }t = t'$).
In Maude, operators are declared with the syntax $\texttt{op } f : s_1 \dots s_n \texttt{ -> } s \texttt{ [} \mathit{attr} \texttt{] .}$, where $s_1, \dots, s_n$ denote domain sorts and $s$ denotes a range sort, and \texttt{attr} is attributes for the operator, such as commutativity or associativity. Maude supports a number of analysis commands. For example, the command \texttt{search [}$n$\texttt{]:} $t$\texttt{ =>* }$t'$\texttt{ such that }$\Psi$\texttt{ .} searches for $n$ states that are reachable from $t$, match the pattern $t'$, and satisfy $\Psi$.

A class declaration $\texttt{class }C \texttt{ | } \mathit{att}_1 \texttt{ : } s_1, \dots, \mathit{att}_n \texttt{ : } s_n$\texttt{ .} declares a class
$C$ with attributes $\mathit{att}_1$ to $\mathit{att}_n$ of sorts $s_1$ to $s_n$. An instance of a class C is represented as a term $\texttt{< }O \texttt{ : } C \texttt{ | } \mathit{att}_1 \texttt{ : } v_1, \dots, \mathit{att}_n \texttt{ : } v_n >$ of sort \texttt{Object}, where $O$ is the object’s identifier, and $v_i$
is the value of each attribute $\mathit{att}_i$. A message is a term of sort \texttt{Msg}. A global system state is a term of sort \texttt{Configuration} that has the structure of a multiset composed of objects and messages, where multiset union is denoted by juxtaposition (empty syntax).

\subsection{\bf Transition Systems and Partial Order Reduction}

A \emph{transition system} $\mathcal{S}$ is defined as a tuple $(S, s_0, T, \AP, L)$ \cite{model-checking-book,Peled2018}, where:
$S$ is the set of states,
$s_0 \in S$ is the initial state,
$T$ represents the set of transitions such that each transition $\alpha \in T$ is a partial function $\alpha : S \to S$,
$\AP$ is a collection of atomic propositions, and
$L : S \rightarrow \power{\AP}$ is a labeling function that associates states with sets of atomic propositions.
A transition $\alpha \in T$ is considered \emph{enabled} in a state $s \in S$ if $\alpha(s)$ is defined. The set of all transitions enabled in state $s$ is denoted by $\enabled{s}$. We often use the notation $s \xrightarrow{\alpha} s'$ to indicate that $\alpha(s) = s'$ for states $s, s' \in S$. Given a rewrite system, a transition system is naturally defined by a mapping from terms to states and rewrites to transitions.


Consider a transition system $\mathcal{S} = (S, s_0, T, \AP, L)$. A transition $\alpha \in T$ is defined as \emph{invisible} if $s \xrightarrow{\alpha} s'$ implies that $L(s) = L(s')$. An \emph{independence relation} $I \subseteq T \times T$ is a symmetric and anti-reflexive relation, such that for any pair of transitions $(\alpha, \beta) \in I$ and a state $s \in S$, where $\alpha, \beta \in \enabled{s}$, the following conditions hold: 
(i) $\alpha \in \enabled{\beta(s)}$ and $\beta \in \enabled{\alpha(s)}$, and 
(ii) $\alpha(\beta(s)) = \beta(\alpha(s))$. 
The complement of $I$, denoted by $D = (T \times T) \setminus I$, is referred to as a dependency relation.

We use partial order reduction via ample sets \cite{Peled2018}. An \emph{ample set} for a state $s \in S$ is a subset of the enabled transitions, represented as $\ample{s} \subseteq \enabled{s}$. A state $s \in S$ is considered \emph{fully expanded} when $\ample{s} = \enabled{s}$. During state space exploration, only the transitions in $\ample{s}$ are examined rather than all transitions in $\enabled{s}$. This process produces a reduced transition system $\hat{\mathcal{S}}$, which maintains behavioral equivalence with the original system when ample sets are chosen correctly.

The following conditions ensure behavioral equivalence (e.g. sttutering bisimulation \cite{model-checking-book}) between the original transition system $\mathcal{S}$ and its reduced version $\hat{\mathcal{S}}$ \cite{Peled2018}: 
(i) $\ample{s} \neq \emptyset$ if and only if $\enabled{s} \neq \emptyset$; 
(ii) any transition that depends on a transition in $\ample{s}$ cannot occur until a transition in $\ample{s}$ occurs first.%
\footnote{For $s \xrightarrow{\beta_1} \cdots \xrightarrow{\beta_n} s_{n} \xrightarrow{\alpha} t$ with $\alpha$ depends on $\ample{s}$, $\beta_i \in \ample{s}$ for some $i \leq n$.};
(iii) if $s$ is not fully expanded, then all transitions in $\ample{s}$ must be invisible; and
(iv) every cycle in the reduced state space $\hat{\mathcal{S}}$ includes at least one fully expanded state.

\subsection{Semantics of PLC ST}

Structured Text (ST) is a textual programming language specified in the IEC 61131-3 standard~\cite{international1993programmable}. It incorporates standard features found in imperative programming languages, including local and global variable assignments, conditional statements, loops, and functions. In addition, ST introduces unique constructs such as function blocks, which are callable objects that maintain an internal state. Functions, function blocks, and programs are collectively known as \emph{program organization units} (POUs).

In short, PLC behavior can be summarized as a repetition of scan cycles. The scan cycle is the continuous loop through which the controller operates to process inputs, execute programs, and update outputs. First, the PLC reads the status of all input devices, such as sensors and switches, and stores these values in memory. Next, it executes the programs based on these inputs. After the program executions, the PLC updates the output devices, such as motors or lights, according to the program results.

Programs are declared using the syntax \texttt{PROGRAM $\mathit{Name}$ ... END\_PROGRAM}, and each program is composed of variable declarations and a code body. Variables are declared with the syntax \texttt{VAR $\mathit{SectionType}$ ... END\_VAR}, where $\mathit{SectionType}$ may be \texttt{GLOBAL}, \texttt{INPUT}, or \texttt{OUTPUT}; if omitted, the variables are considered local. Global variable sections are defined outside of the program. The code body starts after the variable declaration sections.

K~\cite{rosu2010overview} is a rewrite-based semantic framework for defining the semantics of programming languages, grounded in rewriting logic~\cite{unified-tcs}. It has been extensively used to formalize a wide range of languages, including C~\cite{ellison2012executable}, Java~\cite{bogdanas2015k}, JavaScript~\cite{park2015kjs}, PLC Structured Text (ST)\cite{wang2023k,lee2024fm,plcst2022lee}, and AADL\cite{lee2022modeling,hybridsynchaadl}. Several tools can be used to execute and analyze languages using the K framework, including the K tool~\cite{lazar2012executing} and Maude~\cite{maude-manual,cserbuanuctua2010k}.

\begin{figure}[t]
\centering
\tikzstyle{kcell}=[draw, rounded corners,minimum height=1cm, minimum width=1cm]
\begin{tikzpicture}[node distance=0.1cm,font=\footnotesize\smaller]
\node[kcell] (k)  {$\begin{gathered}  \mathbf{k}\\  K  \end{gathered}$};
\node[kcell,right=of k] (env) {$\begin{gathered}  \mathbf{env}\\  \mathit{Id} \mapsto \mathit{Loc}  \end{gathered}$};
\node[kcell,right=of env] (store) {$\begin{gathered}  \mathbf{store}\\  \mathit{Loc} \mapsto \mathit{Val}    \end{gathered}$};
\node[kcell,right=of store] (stack) {$\begin{gathered}  \mathbf{stack}\\  \mathit{List}[(\mathit{Id} \mapsto \mathit{Loc}) \times K]    \end{gathered}$};
\node[kcell,right=of stack] (ctime) {$\begin{gathered}  \mathbf{cycleTime}\\  \mathit{Time}  \end{gathered}$};

\node[kcell,below=of k.south west, anchor=north west] (pouDef) {$\begin{gathered}  \mathbf{pouDef}\\  \mathit{Pid} \mapsto \mathit{VarDecl} \times \mathit{VarDecl}  \times  \mathit{VarDecl} 
        \times \mathit{Code}  \end{gathered}$};
\node[kcell,right=of pouDef] (prog) {$\begin{gathered}  \mathbf{pList}\\  \mathit{List}[\mathit{Pid}]    \end{gathered}$};
\node[kcell,right=of prog] (done) {$\begin{gathered}  \mathbf{done}\\  \mathit{List}[Pid]  \end{gathered}$};
\end{tikzpicture}

\caption{K Cells for PLC ST}
\label{fig:base-config}
\end{figure}
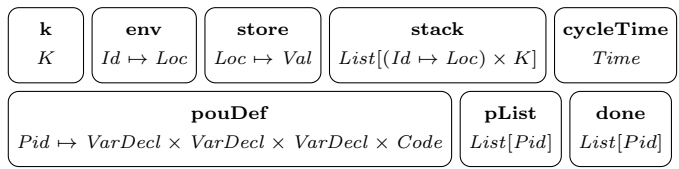

We provide an overview of the basic K semantics for PLC ST~\cite{huang2019kst,plcst2022lee,wang2023k}. \Cref{fig:base-config} illustrates part of the structure of K configurations. $k$ is a cell that includes the following computations, which are tasks to be executed. $\mathit{env}$ and $\mathit{store}$ cells are respectively environment and store. The $\mathit{stack}$ cell contains a call stack, which stores the caller's environment and remaining computations when a function block is invoked. The $\mathit{pouDef}$ cell is a map from POU identifiers to POU declarations, each containing variable declarations (input, output, and local) and code. Lastly, the $\mathit{pList}$ cell holds a list of programs to execute.

\Cref{fig:base-k} presents some of the K rules in the semantics for PLC ST. Due to the modular nature of the K framework~\cite{rosu2010overview,rocsu2014k}, the K rules for standard imperative constructs, such as \texttt{assign} for assignment and \texttt{if-true} and \texttt{if-false} for conditional statements, are nearly identical to those used for other imperative languages, with only minor syntactic adjustments.

%
%

\begin{figure}[t]
\centering
\smaller
\newcommand{\kruleskip}{0.75ex}
\kframe{&
\begin{krule}{assign}{c@{}c@{}c@{}c@{}c@{}}
\langle& x := v& \,\karrow \kdots \rangle_k\
\langle\kdots x \mapsto l \kdots\rangle_\mathit{env}\ 
\langle\kdots l \mapsto &\,\_\,& \,\kdots\rangle_\mathit{store}\ 
\\
\cline{2-2}\cline{4-4}
&\cdot& &v& 
\end{krule}
\\[\kruleskip]&
\begin{krule}{if-true}{c@{}c@{}}
&\texttt{IF $\mathit{true}$ THEN $s_1$ ELSE $s_2$ END\_IF}
\\\cline{2-2}
&s_1 
\end{krule}
\qquad
\begin{krule}{if-false}{c@{}c@{}}
&\texttt{IF $\mathit{false}$ THEN $s_1$ ELSE $s_2$ END\_IF}
\\\cline{2-2}
&s_2
\end{krule}  
}
\caption{Examples of K Rules for PLC ST}
\label{fig:base-k}
\end{figure}


\section{Motivating Example: A Chemical Plant}
\label{sec:motiv}

\begin{figure}[t]
    \centering
    \includegraphics[width=0.7\linewidth]{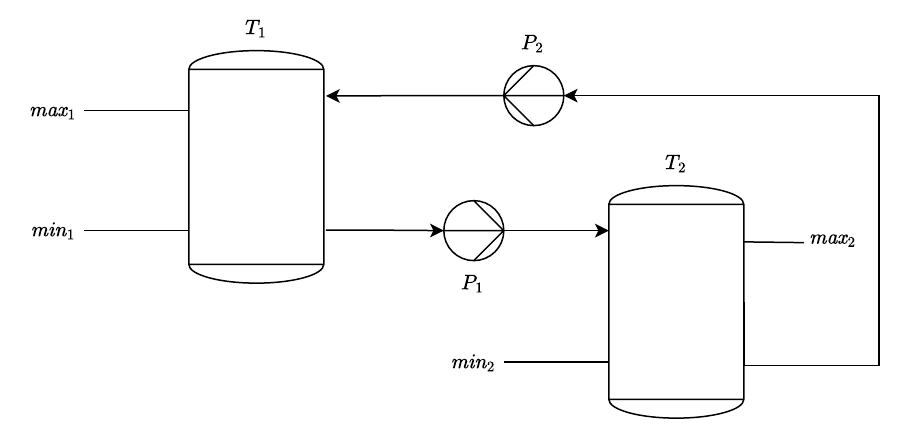}
    \caption{Diagram of a Chemical Plant}
    \label{fig:plant}
\end{figure}


Our motivating example shown in \Cref{fig:plant} is a chemical plant adopted from \cite{nellen2015}. The plant consists of two tanks, $T_1$ and $T_2$, where water is transferred between them by pumps $P_1$ and $P_2$ based on user input, with $P_1$ pumping water from $T_1$ to $T_2$ and $P_2$ pumping water from $T_2$ to $T_1$, while ensuring that both tanks maintain their water levels within specified intervals. This plant is controlled by two PLCs: $\mathit{PLC}_i$ controls pump $P_i$ ($i = 1, 2)$.

While each PLC runs its own program, external behavior is present. When $P_1$ is on, the water level in $T_1$ decreases, while the water level in $T_2$ increases by the same amount. Conversely, when $P_2$ is on, the water level in $T_2$ decreases, and the water level in $T_1$ increases by the same amount. 

PLCs operate cyclically through input, execution, and output stages. In the input stage (sensing), sensor values are copied into input variables. In the output stage (actuation), output variables update physical components, such as motor speed and wheel angle. In the middle, the PLC programs are executed to determine the value of the output values given the sensor inputs. During the execution stage, PLC programs run internally, while externally, physical \emph{flow} occurs as the continuous change of physical attributes. This continuous change is given as a mathematical form, such as ordinary differential equations or solution functions of arbitrary degree.

PLC communication is programmatically controlled by standard function blocks specified in \cite{international1993programmable}, such as \texttt{CONNECT}, \texttt{USEND}, and \texttt{URCV}. The \texttt{CONNECT} function block establishes or disconnects a communication link between two PLCs based on a Boolean trigger. Once the connection is valid, \texttt{USEND} can be used to asynchronously send data from one PLC to another, and \texttt{URCV} to asynchronously receive PLC to access that data.

\begin{figure}[t]
    \centering
    \includegraphics[width=0.7\linewidth]{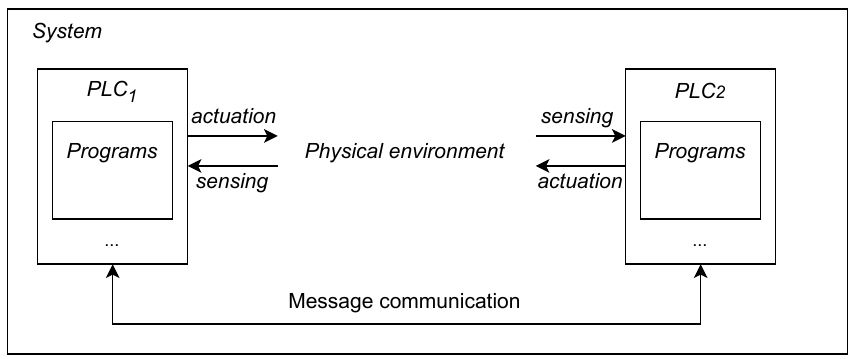}
    \caption{Diagram of an Industrial Control System with PLCs}
    \label{fig:diagram}
\end{figure}

\Cref{fig:diagram} shows a more generalized diagram of the chemical plant. This example includes two PLCs' program executions, a simultaneously flowing environment, and communication between the PLCs. 

Our goal is to develop a model that natively encompasses the whole system's behavior, which is challenging due to the diverse aspects of the system. With this comprehensive semantics, we can run analyses, such as the conformance of the safety properties of the system.
For example, "\textit{$T_1$'s water level is always between 20 and 80}" depend on the interaction of programs, physical environment, and inter-PLC communication. Thus, our semantics is distinctive in its natural ability to handle such properties.

\section{Behavior of PLCs with Physical Environment}
\label{sec:external}

This section explains how we define the semantics which includes PLCs' interaction with the environment. 

\subsection{Overview}
\label{sec:externalOverview}

A PLC system is a set of PLC objects, which exhibit both programmatic and physical behavior.
A PLC object is composed of the PLC's program state, a timer for scan cycles, and a mapping of physical states.
In addition, it includes a flow function that models the evolution of physical quantities over time.

\begin{figure}[t]
    \centering
    \includegraphics[width=0.8\linewidth]{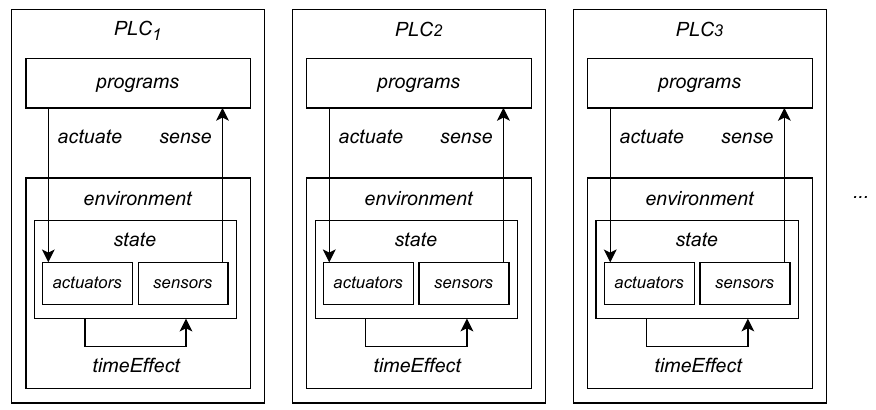}
    \caption{Overview of the System Representation}
    \label{fig:physicalOverview}
\end{figure}

The environment interacts with the programs by sensing and actuation at the beginning of every scan cycle. Sensing and actuation are overwriting values from physical state to program memory back and forth. We define $\mathit{timeEffect}$ that defines the time evolution of the system, which updates the environmental state according to the flow function. We define \texttt{tick} rule, which uses \texttt{timeEffect} to evolve the whole system temporally, and \texttt{start} rule, which starts new scan cycles. Especially, the \texttt{tick} rule captures time progression as discrete transitions, making it suitable for POR discussed in \Cref{sec:reduction}. \Cref{fig:physicalOverview} shows a diagram of our system representation. 

\subsection{Object-oriented Representation}

The following illustrates the class definitions of \texttt{PLCMachine}, which encapsulates the programmatic aspect of PLCs and the external interactions.
\begin{maude}
class PLCMachine | proc: KConfig, timer: Time, state: State, flow: Flow .
\end{maude}
The programmatic part of the \texttt{PLCMachine} is captured in \texttt{proc} attribute, which stores the program configuration in K~\cite{plcst2022lee,rocsu2014k} and describes the state of internal PLC programs. The \texttt{timer} attribute tracks the remaining time in the current scan cycle.
The other attributes account for physical behaviors. Specifically, \texttt{state} is a mapping of actuating/sensing attribute names to their corresponding values, and \texttt{flow} defines the system’s physical dynamics.
\texttt{state} takes the form of comma-separated map entries. 
For example, when the current water level \texttt{waterLevel} is 10 and the pump switch \texttt{pumpSwitch} is zero, the following term represents these attributes in \texttt{state}: 
\begin{maude}
state : waterLevel |-> 10, pumpSwitch |-> 0
\end{maude}
The flow function can be defined as a polynomial equation of arbitrary degree.
In a system where \texttt{waterLevel} decreases by 1 per time unit when \texttt{pumpSwitch} is 1 and remains unchanged when \texttt{pumpSwitch} is 0, the following polynomial equation represents the flow function:
\begin{maude}
flow : waterLevel(t) = waterLevel + (-1) * pumpSwitch * t
\end{maude}
The following represents the whole object with the \texttt{timer} of value 10.
\begin{maude}
< T1 : PLCMachine | proc : app', timer : 10, 
                    state : (waterLevel |-> 10, pumpSwitch |-> 1),
                    flow : waterLevel(t) = waterLevel - pumpSwitch * t >
\end{maude}



\subsection{Semantics of PLC Machines with Physical Attributes}

Two key rewrite rules are defined to specify the PLCs' real-time behavior. The \texttt{start} rule triggers new scan cycles for due objects.
\begin{maude}
crl [start] : {Conf} => {start(Conf)} if Conf =/= start(Conf) .
\end{maude}
The \texttt{tick} rule advances time by an arbitrary duration.
\begin{maude}
crl [tick] : {Conf} => {timeEffect(Conf,T)} in time T if T <= mte(Conf) . 
\end{maude}
It remains to define the \texttt{start}, \texttt{timeEffect}, and \texttt{mte} functions as equations.



A new scan cycle may start when two conditions hold: (1) the PLC’s \texttt{timer} attribute has reached zero, and (2) all programs have completed execution. There are two equations for \texttt{start} operation. The first initiates a new scan cycle for such PLC objects, while the other ensures that objects not due for a new cycle remain unchanged.

The \texttt{start} function then resets the timer and loads the program definitions to begin the next cycle. The specific process of loading and executing programs is assumed to be governed by the semantics of the internal PLC programming language, similar to those in~\cite{plcst2022lee,lee2024fm}. On top of that, \texttt{start} also contains sensing and actuating behavior. \texttt{actuate} and \texttt{sense} are functions actuation and sensing function described in \Cref{sec:motiv}. \texttt{callP} is a K label that loads the programs.
\begin{maude}
var O : Oid .             var KC : KConfig .         var FLOW : Flow .
var ENV : Map{Id, Loc} .  var STORE : Map{Loc, Val} .
vars T TIMER : Time .     var STATE : Map{Id, Val} .
eq start(< O : PLCMachine | timer : 0,  state : STATE,
           proc : k(.) cycleTime(T) env(ENV) store(STORE) KC >) 
= < O : PLCMachine | timer : T, state : actuate(ENV, STORE, STATE),
                     proc : k(callP) cycleTime(T) env(ENV)
                     store(sense(ENV, STORE, STATE)) KC > .
eq start(< O : PLCMachine | timer : T >) = < O : PLCMachine | > [owise] .
\end{maude}

To model the passage of time, the system defines a \emph{time effect function}. The following equation describes the \emph{time effect} of duration \texttt{T} for a single \texttt{PLCMachine} object, provided that \texttt{T} does not exceed the object's current \texttt{timer} value \texttt{TIMER}. The function \texttt{x monus y} is defined as \( \max(\texttt{x - y}, 0) \), ensuring that time never progresses past zero. \texttt{PLCMachine} also must reflect the physical behavior by updating the \texttt{state} attribute according to \texttt{flow} attribute. The \texttt{eval} function in \texttt{state} attribute evaluates the result of the physical flow of a given duration.

\begin{maude}
eq timeEffect(
< O : PLCMachine | timer: TIMER, flow: FLOW, state: STATE >, T) = 
< O : PLCMachine | timer: TIMER monus T, state: eval(FLOW, STATE, T) > .
\end{maude}

The \texttt{mte} (\emph{maximal time elapse}) function determines the maximum duration the system can advance before another transition (such as a new scan cycle) must occur. For a single \texttt{PLCMachine} object, the \texttt{mte} value is simply the \texttt{timer} value, ensuring that time does not surpass the scan cycle boundary.

\begin{maude}
eq mte(< O : PLCMachine | timer : TIMER >) = TIMER .
\end{maude}

\subsection{Example}
\label{sec:environment-example}
We revisit the motivating example introduced in \Cref{sec:motiv}, but without communication.
The whole system including two PLCs and their physical environment must be handled at once, but we only explain in the view of one PLC machine for the sake of simplicity. Consider the state of the water tank $T_1$ represented by the following object:
\begin{maude}    
< T1 : PLCMachine | proc : app, timer : 10, 
                    state : waterLevel |-> 10, pumpSwitch |-> 0,
                    flow : F(t) = waterLevel + (-1) * pumpSwitch * t >
\end{maude}
After applying \texttt{tick} rule of duration 10 will result in the following state.
\begin{maude}    
< T1 : PLCMachine | proc : app, timer : 0, 
                    state : waterLevel |-> 10, pumpSwitch |-> 0,
                    flow : F(t) = waterLevel + (-1) * pumpSwitch * t >
\end{maude}
There is only one sensor attribute \texttt{waterLevel} and one actuation attribute \texttt{pumpSwitch}. Thus, the flow function $\texttt{F}$ is a scalar function for \texttt{waterLevel}.\footnote{as mentioned in \Cref{sec:motiv}, the flow is not restrained to a linear function.} Since \texttt{pumpSwitch} has value 0 in \texttt{state} attribute, the flow function \texttt{F} returns the current \texttt{waterLevel}. This means there is no change in $T_1$'s water level because the pump is off. Note that the timer value is reduced from 10 to 0.
The \texttt{app} is a K configuration including the program code, store, and environment. The following code is loaded in $T_1$.

\noindent
\begin{minipage}{0.48\linewidth}
\begin{plc}
PROGRAM T1
VAR_INPUT
  waterLevel : REAL;
END_VAR
VAR_OUTPUT
  pumpSwitch : INT;
END_VAR
VAR
  input : BOOL;
\end{plc}
\end{minipage}
\begin{minipage}{0.5\linewidth}
\begin{plc}
  ...
END_VAR

  IF input THEN
    pumpSwitch := 1;
  ELSE
    pumpSwitch := 0;
  END_IF;
END_PROGRAM
\end{plc}
\end{minipage}

The sensing attribute \texttt{waterLevel} and the actuation attribute \texttt{pumpSwitch} are declared as input and output variable with the same names. \texttt{input} is a Boolean value that is directly linked to user input. Depending on the user input, \texttt{pumpSwitch} is 1 or 0. Initially, \texttt{pumpSwitch} holds value 0, and let us assume that the user gives \texttt{1} in this scan cycle. After finishing the execution of this program, \texttt{pumpSwitch} in \texttt{state} still holds 0. Let \texttt{app'} represent the final program state after executing this program.
\begin{maude}
< T1 : PLCMachine | proc : app', timer : 0, 
                    state : waterLevel |-> 10, pumpSwitch |-> 0,
            flow : waterLevel(t) = waterLevel + (-1) * pumpSwitch * t >
\end{maude}

The rewrite rule \texttt{start} may apply when there is a component with \texttt{timer} 0. Assume the $T_2$'s timer is also 0. When \texttt{start} applies, the system configuration goes through actuation, sensing, and reloading programs. It also resets the PLC’s \texttt{timer} to the specified cycle time. Since \texttt{waterLevel} has not been changed, sensing makes no change. However, actuation updates the value of \texttt{pumpSwitch} from 0 to 1 in \texttt{state} attribute. The following is the resulting object configuration. \texttt{app''} represents the program state where programs are reloaded in \texttt{app'}.
\begin{maude}
< T1 : PLCMachine | proc : app'', timer : 10,
                    state : waterLevel |-> 10, pumpSwitch |-> 1,
            flow : waterLevel(t) = waterLevel + (-1) * pumpSwitch * t >
\end{maude}

Since \texttt{timer} is set to 10, we may apply \texttt{tick} rule of duration up to 10. When 6 time unit has elapsed, since \texttt{pumpSwitch} in \texttt{state} has value 1, the water level becomes 4 according to the flow function.
\begin{maude}
< T1 : PLCMachine | proc : app'', timer : 4, 
                    state : waterLevel |-> 4,  pumpSwitch |-> 1,
            flow : waterLevel(t) = waterLevel + (-1) * pumpSwitch * t >
\end{maude}

\section{Semantics of Communication}
\label{sec:communication}

In this section, we describe our formal semantics of communication that conforms to IEC 61131~\cite{international1993programmable} standard. We focus on connection establishment and asynchronous send/receive functionality. The behaviors of such function blocks, \texttt{CONNECT}, \texttt{USEND}, and \texttt{URCV} are described in \Cref{sec:motiv}.

\subsection{Overview}

We define the behavior of communication function blocks in PLC ST in the form of user-defined function blocks. These function blocks are stateful and operate across multiple steps, so it is inappropriate to model them as single transitions. Specifically, a single run of a function block call involves conditional branching, flag settings, and atomic operations such as connection establishment, data transmission, and data reception. To enable POR, we carefully identify atomic operations and encode communication behavior using function blocks constructed from these operations.

We formulate the $\mathit{Conn}$ class representing a connection, which collects sent messages so that the receivers can retrieve them. A $\mathit{Conn}$ object may exist for every pair of PLCs in the system. It stores the messages in their transmission and the network delay constraint.


The handling of connection establishment, message transmission, and message reception necessitates the introduction of new semantic components. To address this, we define atomic operations within the \texttt{proc} attribute to structure these operations: \texttt{connectRequest} for initiating connection establishment, \texttt{isConnected} for checking connection establishment, \texttt{sendData} for transmitting data, and \texttt{rcvData} for receiving data.




We introduce a time assertion annotation to express timing constraints on the execution of individual statements within a scan cycle. This annotation specifies that a particular statement must be executed after a given time range has elapsed since the beginning of the scan cycle. \footnote{Extensive research has been dedicated to the analysis of worst-case execution time (WCET). The survey in \cite{wcetsurvey} offers a detailed overview of existing techniques and tools addressing the WCET challenge, while \cite{wcrtiec} specifically explores WCET estimation for function block calls in IEC 61499 systems.

Beyond computation, communication delays can also be bounded in real-time systems through suitable architectural designs. For example, \cite{ttaverif} outlines a formal verification approach for time-triggered architectures that incorporate communication protocols, and \cite{tte} proposes a time-triggered Ethernet (TTE) design to guarantee predictable network latency. Together, these works provide a solid foundation for the correctness of our time-related annotations.}
Although our formal semantics is designed to accommodate arbitrary execution times of statements within a scan cycle, this can lead to overly conservative or unrealistic behavior during analysis. \footnote{Note that we can choose not to use time annotations and still have sound results. By using time assertion annotations, we can constrain the space of possible executions to those that align more closely with real-world implementations, such as fixed task scheduling or known latencies.
We assume that certain system parameters—such as worst-case execution times and maximum network delays—are known to PLC developers, which is a common practice in real-world settings.} 
Time assertion annotation takes \texttt{//assertTime(}$\mathit{min}, \mathit{max}$\texttt{)} form.
\begin{plc}
//assertTime(50, 100)
send(TRUE, con, rcv', data);
...
//assertTime(200, 250)
rcv(TRUE, con, send');
\end{plc}
\texttt{con} is a connection object, \texttt{send} and \texttt{rcv} are respectively \texttt{USEND}, \texttt{URCV} function block instances. \texttt{send'} and \texttt{rcv'} are the function block instances of the other PLC object connected through \texttt{con}.
The first assertion ensures that the invocation \texttt{send} is executed during 50 to 100 ms after the beginning of the scan cycle. Similarly, the second assertion restricts the invocation of \texttt{rcv} to take place from 200 to 250 ms in each cycle. 

For \texttt{send} function blocks, user may specify the minimum and maximum time for the sent message to be available in the receiver.
The message delay annotation is represented \texttt{//delay(P1, P2, m, M)}, where \texttt{P1} and \texttt{P2} are object ids, \texttt{m} is a minimum message delay, and \texttt{M} is a maximum message delay. The following annotation makes the \texttt{data} arrive in 10 to 20 ms after the message creation.
\begin{plc}
//delay(P1, P2, 10, 20)
send(TRUE, con, rcv', data);
\end{plc}

%


\subsection{Encoding of Communication Function Blocks}
\label{sec:commencoding}

\Cref{fig:connect-code} is the function block definition of \texttt{CONNECT}. The function block has two inputs: \texttt{ENC}, a Boolean that triggers the connection process, and \texttt{PARTNER}, a string identifying the target device. It produces four outputs: \texttt{VALID}, indicating a successful connection; \texttt{ERROR}, set to \texttt{TRUE} if a failure occurs; \texttt{STATUS}, an integer representing the connection state (0 for success, 1 for failure); and \texttt{ID}, storing the connected partner's identifier. The connection process starts when \texttt{ENC} is \texttt{TRUE}, initiating a request to \texttt{PARTNER}. If the connection is valid and \texttt{ENC} switches to \texttt{FALSE}, disconnection follows. The block continuously checks the connection status using \texttt{isConnected(PARTNER)}, updating outputs accordingly: if connected, \texttt{VALID} is \texttt{TRUE}, \texttt{ERROR} is \texttt{FALSE}, \texttt{STATUS} is 0, and \texttt{ID} is updated; otherwise, \texttt{VALID} is \texttt{FALSE}, \texttt{ERROR} is \texttt{TRUE}, and \texttt{STATUS} is 1.

\Cref{fig:usend-code} shows the ST definition of \texttt{USEND}.
It accepts input variables such as \texttt{REQ}, a boolean that triggers the data transmission request, \texttt{COMMCHANNEL}, a string representing the communication channel, \texttt{RID}, a string denoting the recipient's identifier, and \texttt{DATA}, an \texttt{ANY} type for the data being sent. The block outputs three variables: \texttt{DONE}, a boolean indicating successful completion of the transmission, \texttt{ERROR}, a boolean flagging any issues during transmission, and \texttt{STATUS}, a \texttt{DINT} variable representing the current status code. The internal logic starts by checking if the communication channel \texttt{COMMCHANNEL} is connected using the \texttt{isConnected} function. If the channel is not connected, \texttt{ERROR} is set to \texttt{TRUE}, \texttt{DONE} remains \texttt{FALSE}, and \texttt{STATUS} is updated to \texttt{1}. If the channel is connected, the block proceeds to call the \texttt{sendData} function, passing \texttt{COMMCHANNEL}, \texttt{THIS} (the current block identifier), \texttt{RID}, and \texttt{DATA} as parameters. The result of the \texttt{sendData} call is stored in \texttt{RESULT}. If \texttt{RESULT} is \texttt{TRUE}, indicating successful transmission, \texttt{DONE} is set to \texttt{TRUE}, \texttt{ERROR} is reset to \texttt{FALSE}, and \texttt{STATUS} is set to \texttt{0}. Otherwise, the function flags an error and updates the status accordingly.

Lastly, \Cref{fig:urcv-code} shows the ST definition of \texttt{URCV}.
It takes input variables such as \texttt{ENR}, a boolean that enables the data reception process, \texttt{ID}, a string identifying the communication channel, and \texttt{RID}, a string representing the sender's identifier. The block provides output variables: \texttt{NDR}, a boolean flag indicating that new data has been received, \texttt{ERROR}, a boolean flagging any errors during the reception, \texttt{STATUS}, a \texttt{DINT} representing the current status, and \texttt{DATA}, which stores the received data. Internally, the function first resets \texttt{NDR}, \texttt{ERROR}, and \texttt{STATUS} to their default values when \texttt{NDR} is \texttt{TRUE}, effectively clearing any previous transmission states. If the communication channel identified by \texttt{ID} is not connected, \texttt{ERROR} is set to \texttt{TRUE}, and \texttt{STATUS} is updated to \texttt{1}. The function then calls \texttt{rcvData} with the channel ID \texttt{ID}, sender ID \texttt{RID}, and the current block identifier \texttt{THIS}. If the reception is successful (i.e., \texttt{RESULT} is not equal to \texttt{rcvError}), \texttt{NDR} is set to \texttt{TRUE}, \texttt{ERROR} is reset to \texttt{FALSE}, \texttt{STATUS} is set to \texttt{0}, and the received data is stored in the \texttt{DATA} variable. If an error occurs during reception, \texttt{NDR} remains \texttt{FALSE}, \texttt{ERROR} is set to \texttt{TRUE}, and \texttt{STATUS} is updated to \texttt{1}.

\begin{figure}[t]
\begin{minipage}{0.48\linewidth}
\begin{plc}
FUNCTION_BLOCK CONNECT
VAR_INPUT
ENC : BOOL; PARTNER : STRING;
END_VAR
VAR_OUTPUT
VALID : BOOL := FALSE ;
ERROR : BOOL := FALSE ;
STATUS : DINT := 0 ;
ID : STRING;
END_VAR
  IF ENC = TRUE THEN
    connectRequest(PARTNER);
  END_IF ;
\end{plc}
\end{minipage}
\hfill
\begin{minipage}{0.5\linewidth}
\begin{plc}
  IF VALID AND NOT ENC THEN
    disconnect(PARTNER);
  END_IF ;    
  IF isConnected(PARTNER) THEN
    VALID := TRUE ;    ERROR := FALSE ;
    STATUS := 0;
    ID := PARTNER ;
  ELSE
    VALID := FALSE ;
    ERROR := TRUE ;
    STATUS := 1;
  END_IF ;
END_FUNCTION_BLOCK
\end{plc}
\end{minipage}
    \caption{Function Block Encoding of \texttt{CONNECT}}
    \label{fig:connect-code}
\end{figure}

\begin{figure}[t]
\begin{minipage}{0.48\linewidth}
\begin{plc}
FUNCTION_BLOCK USEND
VAR_INPUT
REQ : BOOL;         COMM : STRING;
RID : STRING;       DATA : ANY;
END_VAR
VAR_OUTPUT
DONE : BOOL := FALSE ;
ERROR : BOOL := FALSE ;
STATUS : DINT := 0 ;
END_VAR
VAR
THIS : STRING ;
RESULT : BOOL := FALSE ;
END_VAR
  IF RESULT THEN
    DONE := FALSE ; ERROR := FALSE ;
\end{plc}
\end{minipage}
\hfill
\begin{minipage}{0.5\linewidth}
\begin{plc}
    STATUS := 0 ;
  END_IF ;
  IF isConnected(COMM) = FALSE THEN
    DONE := FALSE ;
    ERROR := TRUE ;
    STATUS := 1 ;
  END_IF ;
  THIS := thisBlock ;
  RESULT := 
    sendData(COMM, THIS, RID, DATA);
  IF RESULT THEN
    DONE := TRUE ;
    ERROR := FALSE ;
    STATUS := 0 ;
  END_IF ;
END_FUNCTION_BLOCK
\end{plc}
\end{minipage}
    \caption{Function Block Encoding of \texttt{USEND}}
    \label{fig:usend-code}
\end{figure}

\begin{figure}[t]
\begin{minipage}{0.48\linewidth}
\begin{plc}
FUNCTION_BLOCK URCV
VAR_INPUT
ENR : BOOL;  ID : STRING;
RID : STRING;
END_VAR
VAR_OUTPUT
NDR : BOOL := FALSE ;
ERROR : BOOL := FALSE ;
STATUS : DINT := 0 ;
DATA : ANY;
END_VAR
VAR
THIS : STRING ; RESULT : ANY ;
END_VAR
  IF NDR THEN
    NDR := FALSE ;   ERROR := FALSE ;
    STATUS := 0 ;
  END_IF ;
\end{plc}
\end{minipage}
\hfill
\begin{minipage}{0.5\linewidth}
\begin{plc}
  RETURN ;
  IF isConnected(ID) = FALSE THEN
    ERROR := TRUE ;
    STATUS := 1 ;
  END_IF ;
  THIS := thisBlock ;
  RESULT := rcvData(ID, RID, THIS) ;
  IF RESULT <> rcvError THEN
    NDR := TRUE ;
    ERROR := FALSE ;
    STATUS := 0 ;
    DATA := RESULT ;
  ELSE
    NDR := FALSE ;
    ERROR := TRUE ;
    STATUS := 1 ;
  END_IF ;
END_FUNCTION_BLOCK
\end{plc}
\end{minipage}
    \caption{Function Block Encoding of \texttt{URCV}}
    \label{fig:urcv-code}
\end{figure}

\subsection{Class Definition and Semantics of Communication between PLCs}

All communication behaviors are captured by functions that modify or refer to \texttt{Conn} objects. The following is the class definition of \texttt{Conn}, the constructor for its identifiers, and the message constructor of messages.
\begin{maude}
class Conn | validity: Bool, buffer: Set{Msg}, delay: Time * Time .
op conn : Oid Oid -> Oid [ctor comm] .
op m : Oid Oid Id Id Any Time Time -> Msg [ctor] .
\end{maude}
\texttt{validity} is an attribute that denotes if the connection is successfully established and functioning. The \texttt{buffer} stores the sent message along with information about the source and destination function block instances. \texttt{delay} contains the minimum and maximum message delay specified by the delay annotation. The identifier of the \(\mathit{Conn}\) between PLCs \(O\) and \(O'\) is \(\texttt{conn(} O \texttt{, } O' \texttt{)}\).\footnote{To avoid redundant definitions for the same pair of PLCs (e.g., \(\texttt{conn(} O \texttt{, } O' \texttt{)}\) and \(\texttt{conn(} O' \texttt{, } O \texttt{)}\)), the identifier constructor is defined using the \texttt{comm} axiom.}

A message is composed of the sender's ID, the receiver's ID, the identifiers of the sending and receiving function block instances, the transmitted data, and two message delay timers. Each message is assigned delay timers upon creation according to the delay annotation. The \texttt{timeEffect} function, when applied to a \texttt{Conn} object, decreases the delay timers of all messages by the elapsed time. Once a message's minimum delay timer reaches zero, it becomes eligible for reception by the target machine. When a message's maximum delay timer is zero, it must be accepted before time elapses. The \texttt{decreaseTimer} function iterates through messages in the buffer, reducing the delay timers by the specified amount. 
\begin{maude}
vars C O O' : Oid .   var BUFFER : Set{Msg} .
var K : K .           var KC : KConfig .     vars L U : Time .
var DATA : Val .      var B : Bool .         vars T T2 : Time .
vars MIN MAX MIN' MAX' : Time .              vars SFBID RFBID : Id .

eq timeEffect(< C : Conn | buffer : BUFFER >, T)
 = < C : Conn | buffer : decreaseTimer(BUFFER, T) > .
eq decreaseTimer(empty, T) = empty . 
eq decreaseTimer((m(O, O', SFBID, RFBID, DATA, MIN, MAX) BUFFER), T)
 = m(O, O', SFBID, RFBID, DATA, MIN monus T, MAX monus T) 
   decreaseTimer(BUFFER) .
\end{maude}

The maximal time elapse of \texttt{Conn} object is the minimum value of all the messages' maximum delay.
\begin{maude}
op minMaxDelay : Set{Msg} ~> Time .
eq mte(< C : Conn | buffer : BUFFER >) = minMaxDelay(BUFFER) .
eq minMaxDelay(empty) = infinity .
eq minMaxDelay((m(O, O', SSFBID, RFBID, DATA, MIN, MAX), BUFFER))
 = min(MAX, minMaxDelay(BUFFER)) .
\end{maude}

The following describes the atomic operations for communication, including \texttt{connectRequest}, \texttt{isConnected}, \texttt{sendData}, and \texttt{rcvData}. Before a message passing, a connection must be established first.
\texttt{connectRequest} takes a PLC identifier and opens a connection to that PLC. Its output includes the connection's name and its validity. \texttt{conSucc} and \texttt{conFail} are rules that respectively define the success or failure cases of the requested connection.
\begin{maude}
rl [conSucc] :
   < O : PLC | proc : k(connectRequest(O') ~> K) KC >
   < conn(O, O') : Conn | validity : false > 
=> < O : PLC | proc : k(K) KC >
   < conn(O, O') : Conn | > .
   
rl [conFail] :
   < O : PLC | proc : k(connectRequest(O') ~> K) KC >
   < connection(O, O') : Channel | validity : false >
=> < O : PLC | proc : k(K) KC >
   < connection(O, O') : Channel | > .
\end{maude}
\texttt{isConnected} is a function that checks if the desired connection is successfully established. It takes a PLC name and returns either \texttt{TRUE} or \texttt{FALSE} depending on the presence of the established connection to that PLC. 
\begin{maude}
rl [conCheck]:
   < O : PLC | proc : k(isConnected(O') ~> K) KC >
   < conn(O, O') : Conn | validity : B >
=> < O : PLC | proc : k(if B then TRUE else FALSE fi ~> K) KC >
   < conn(O, O') : Conn | > .
\end{maude}
\texttt{sendData} sends data when given the sender's and receiver's PLC identifiers and the function block instance identifiers of \texttt{USEND} and \texttt{URCV}. The message delay for the message to be sent is in the \texttt{delay} attribute. The \texttt{sendData} and \texttt{sendDataFail} rules, which respectively succeed and fail to dispatch the data.
\begin{maude}
rl [sendData] :
   < O : PLC | proc : k(sendData(O', SFBID, RFBID, DATA) ~> K) KC >
   < conn(O, O') : Conn | validity : true, buffer : BUFFER, 
                          delay : (MIN, MAX) >
=> < O : PLC | proc : k(TRUE ~> K) KC >
   < conn(O, O') : Conn | buffer : insert(BUFFER, 
                            m(O, O', SFBID, RFBID, DATA, MIN, MAX)) > .

rl [sendDataFail] :
   < O : PLC | proc : k(sendData(O', SFBID, RFBID, DATA) ~> K) KC >
   < conn(O, O') : Conn | validity : false, buffer : BUFFER >
=> < O : PLC | proc : k(FALSE ~> K) KC >
   < conn(O, O') : Conn | > .
\end{maude}
The function \texttt{rcvData} is a function that accepts the data if available in the corresponding buffer of the connection. The series of rules in the following describes success (\texttt{rcvData}), failure due to lost connection (\texttt{rcvFail}), or accepting no message (\texttt{rcvNo}).
\begin{maude}
crl [rcvData] :
    < O : PLC | proc : k(rcvData(O', SFBID, RFBID) ~> K) KC >
    < conn(O, O') : Conn | validity : true, 
            buffer : (BUFFER m(O', O, SFBID, RFBID, DATA, MIN, MAX)) >
 => < O : PLC | proc : k(DATA ~> K) KC >
    < conn(O, O') : Conn | buffer : BUFFER >
 if MIN == 0 .
 
 rl [rcvFail] :
    < O : PLC | proc : k(rcvData(O', SFBID, RFBID) ~> K) KC >
    < conn(O, O') : Conn | validity : false, buffer : BUFFER >
 => < O : PLC | proc : k(rcvError ~> K) KC >
    < conn(O, O') : Conn | > .

crl [rcvNo] :
   < O : PLC | proc : k(rcvData(O', SFBID, RFBID) ~> K) KC >
   < conn(O, O') : Conn | validity : true, buffer : BUFFER >
=> < O : PLC | proc : k(rcvError ~> K) KC >
   < conn(O, O') : Conn | > if not checkMsg(BUFFER, O', SFBID, RFBID) .
\end{maude}

The time annotation has the following constructors.
\begin{maude}
sort Annotation .       subsort Annotation < K .
op //assertTime : Time Time -> Annotation [ctor] .
op //delay : Oid Oid Time Time -> Annotation [ctor] .
\end{maude}

The rules for time annotations are as follows. The rule for time assertion checks that the elapsed time from the beginning of the current cycle \texttt{T2 - T} is in the assertion range $[\texttt{L}, \texttt{U}]$. The rule for message delay replaces the original content of \texttt{delay} attribute with the given values.
\begin{maude}
crl < O : PLCMachine | timer : T, proc : k(//assertTime(L, U) ~> K) 
                       cycleTime(T2) KC, ATTRS >
 => < O : PLCMachine | timer : T, proc : k(K) cycleTime(T2) KC, ATTRS >
  if L <= T2 - T and U >= T2 - T .

 rl < O : PLCMachine | proc : k(//delay(O, O', MAX, MIN) ~> K), ATTRS >
    < conn(O, O') : Conn | delay : (MAX', MIN') >                   
 => < O : PLCMachine | proc : k(K), ATTRS >
    < conn(O, O') : Conn | delay : (MAX, MIN) > .
\end{maude}


\subsection{Example}
\label{sec:comm-example}
Consider the example in \Cref{sec:environment-example}, but this time with communication. The following is the programs installed in $T_1$ and $T_2$.

\noindent
\begin{minipage}{0.48\textwidth}
\begin{plc}
PROGRAM T1
VAR_INPUT
  waterLevel : REAL;
END_VAR
VAR_OUTPUT
  pumpSwitch : INT;
END_VAR
VAR
  input : INT;         comm : CONNECT;
  send : USEND;       rcv : URCV;
  sig_in : INT;       sig_out : INT; ...
END_VAR
  comm(TRUE , "T2");
  IF NOT comm.VALID THEN
    RETURN;
  END_IF;
  sig_out := input;
  send(TRUE, "T2", "rcv", sig_out);
  rcv(TRUE, "T2", "send");
  sig_in := rcv.DATA;
  pump_switch := sig_out - sig_in;
END_PROGRAM
\end{plc}    
\end{minipage}
\hfill
\begin{minipage}{0.5\textwidth}
\begin{plc}
PROGRAM T2
VAR_INPUT
  waterLevel : REAL;
END_VAR
VAR_OUTPUT
  pumpSwitch : INT;
END_VAR
VAR
  input : INT;     comm : CONNECT;
  send : USEND;   rcv : URCV;
  sig_in : INT;   sig_out : INT; ...
END_VAR
  comm(TRUE , "T1");
  IF NOT comm.VALID THEN
    RETURN;
  END_IF;
  sig_out := input;
  send(TRUE, "T1", "rcv", sig_out);
  rcv(TRUE, "T1", "send");
  sig_in := rcv.DATA;
  pump_switch := sig_out - sig_in;
END_PROGRAM
\end{plc}    
\end{minipage}

The process begins with an attempt to establish a connection using the \texttt{comm(TRUE, "T2")} function, where \texttt{"T2"} represents $T_2$.
After initiating communication, the program verifies whether the connection was successfully established by checking the \texttt{comm.VALID} flag. If the connection is invalid, the program terminates early using the \texttt{RETURN} statement. Once communication is validated, the program assigns the value of \texttt{input} to \texttt{signal\_out}, preparing the data for transmission. The \texttt{send(TRUE, "T2", "rcv", signal\_out)} function sends \texttt{signal\_out} to $T_2$'s receiving function block instance \texttt{rcv}. The program then executes \texttt{rcv(TRUE, "T2", "send")}, which retrieves data from $T_2$ that was sent by a function block instance \texttt{"send"}. The received data in \texttt{rcv.DATA} is assigned to \texttt{signal\_in}. Finally, the difference between \texttt{signal\_out} (the sent data) and \texttt{signal\_in} (the received data) is computed and stored in \texttt{pump\_switch}. This ensures that when both tanks receive positive input (+1) from users, they do not activate their pumps simultaneously, preventing meaningless water exchange.

Now consider the following state where the first scan cycle has begun, invoked \texttt{comm(...)}, and is ready to execute \texttt{connectRequest(T2)}. The message delay time is in the default setting, which is 10 to 20 ms.
\begin{maude}
< T1 : PLCMachine | proc : k(connectRequest(T2) ~> ...) ..., timer : 10,
                    state : waterLevel |-> 10, pumpSwitch |-> 0,
                    flow : F(t) = waterLevel + (-1) * pumpSwitch * t >
< T2 : PLCMachine | proc : k(...) ..., timer : 10, 
                    state : waterLevel |-> 10, pumpSwitch |-> 0,
                    flow : F(t) = waterLevel + (-1) * pumpSwitch * t >
< conn(T1, T2) : Conn | 
                 validity : false, buffer : empty, delay : (10, 20) >
\end{maude}
\texttt{con-succ} rule is applied to set the connection object's validity to true.
\begin{maude}
< T1 : PLCMachine | proc : k(...) ..., timer : 10, 
                    state : waterLevel |-> 10, pumpSwitch |-> 0,
                    flow : F(t) = waterLevel + (-1) * pumpSwitch * t >
< T2 : PLCMachine | proc : k(...) ..., timer : 10, 
                    state : waterLevel |-> 10, pumpSwitch |-> 0,
                    flow : F(t) = waterLevel + (-1) * pumpSwitch * t >
< conn(T1, T2) : Conn | 
                  validity : true, buffer : empty, delay : (10, 20) >
\end{maude}
After 3 time units, $T_1$ is ready to send its signal inside the context of \texttt{send}. The signal is given as 1 by the user.
\begin{maude}
< T1 : PLCMachine | proc : k(sendData(T2, "send", "rcv", 1) ~> ...) ..., 
                    timer : 7, 
                    state : waterLevel |-> 10, pumpSwitch |-> 0,
                    flow : F(t) = waterLevel + (-1) * pumpSwitch * t >
< T2 : PLCMachine | proc : k(...) ..., 
                    timer : 7, 
                    state : waterLevel |-> 10, pumpSwitch |-> 0,
                    flow : F(t) = waterLevel + (-1) * pumpSwitch * t >
< conn(T1, T2) : Conn | 
                  validity : true, buffer : empty, delay : (10, 20) >
\end{maude}
Rewrite rule \texttt{sendData} applies and inserts the intended message to the \texttt{buffer} attribute in the connection object.
\begin{maude}
< T1 : PLCMachine | proc : k(...) ..., timer : 7, 
                    state : waterLevel |-> 10, pumpSwitch |-> 0,
                    flow : F(t) = waterLevel + (-1) * pumpSwitch * t >
< T2 : PLCMachine | proc : k(...) ..., timer : 7, 
                    state : waterLevel |-> 10, pumpSwitch |-> 0,
                    flow : F(t) = waterLevel + (-1) * pumpSwitch * t >
< conn(T1,T2) : Conn | validity : true, 
          buffer : (T1, T2, "send", "rcv", 10, 20), delay : (10, 20) >
\end{maude}
After 10 milliseconds, during $T_2$'s invocation of the function block instance \texttt{rcv}, it is time to process \texttt{rcv} function.  
\begin{maude}
< T1 : PLCMachine | proc : k(...), timer : 5, 
                    state : waterLevel |-> 10, pumpSwitch |-> 0,
                    flow : F(t) = waterLevel + (-1) * pumpSwitch * t >
< T2 : PLCMachine | proc : k(rcv(T1, "send", "rcv") ~> ...), timer : 5, 
                    state : waterLevel |-> 10, pumpSwitch |-> 0,
                    flow : F(t) = waterLevel + (-1) * pumpSwitch * t >
< conn(T1,T2) : Conn | validity : true, 
          buffer : (T1, T2, "send", "rcv", 0, 10),  delay : (10, 20) >
\end{maude}
Applying \texttt{rcv-data} transfers the data from the connection object to $T_2$'s K configuration.
\begin{maude}
< T1 : PLCMachine | proc : k(...) ..., timer : 5, 
                    state : waterLevel |-> 10, pumpSwitch |-> 0,
                    flow : F(t) = waterLevel + (-1) * pumpSwitch * t >
< T2 : PLCMachine | proc : k(1 ~> ...) ..., timer : 5, 
                    state : waterLevel |-> 10, pumpSwitch |-> 0,
                    flow : F(t) = waterLevel + (-1) * pumpSwitch * t >
< conn(T1,T2) : Conn | validity : true, buffer : empty, delay : (10, 20)>
\end{maude}


\section{Partial Order Reduction}
\label{sec:reduction}

\begin{figure}[t]
    \centering
    \includegraphics[width=1.0\linewidth]{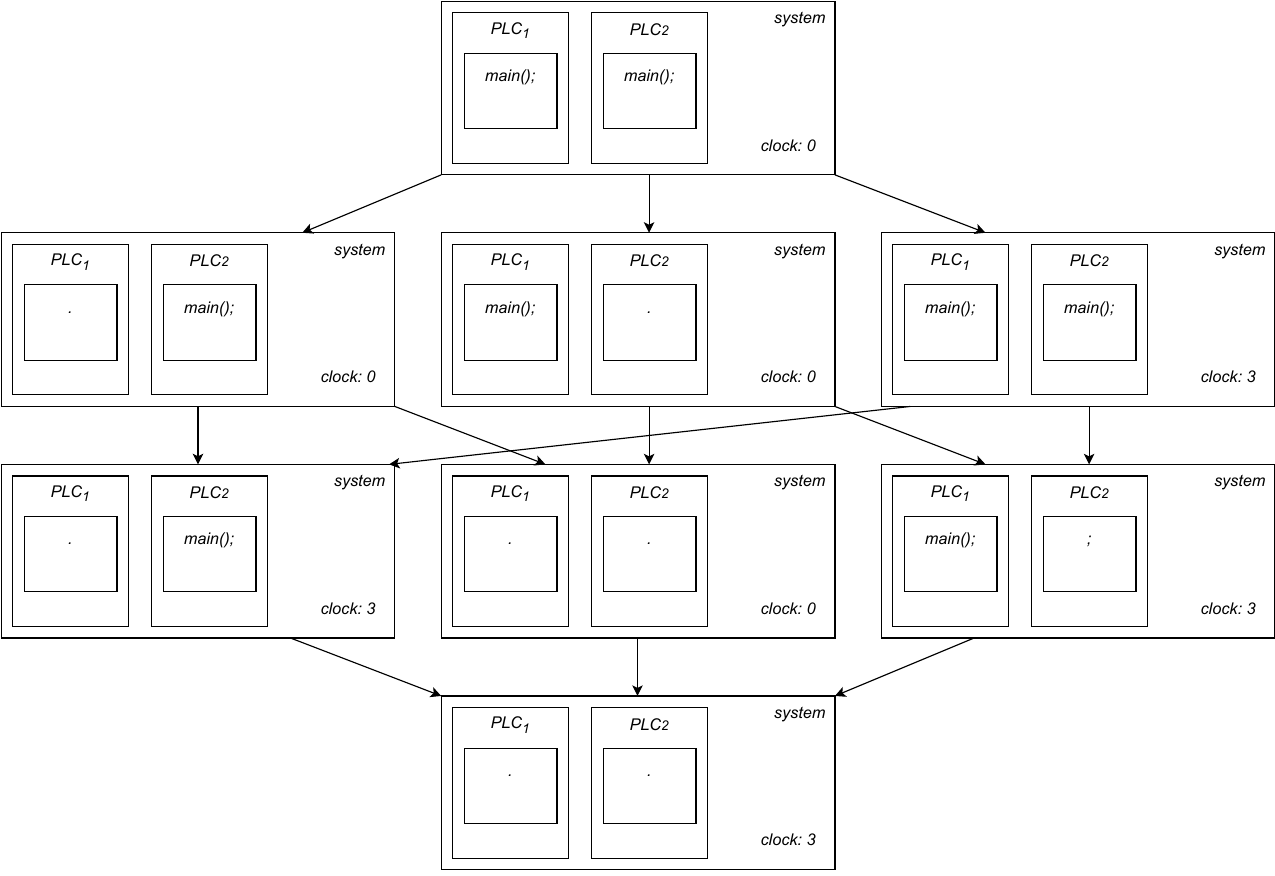}
    \caption{State transition diagram without POR}
    \label{fig:portransition}
\end{figure}

Despite the conciseness of our semantics, state explosion remains an issue due to redundant state exploration. Consider the state transition diagram in \Cref{fig:portransition}. The top state is where two PLCs are ready to execute their own $\mathit{main}$ function, and the system clock is 0. The system can take one of the three choices: to execute $\mathit{PLC}_1$'s program, to execute $\mathit{PLC}_2$'s program, and to elapse time by 3 unit. Similarly, each of the resulting three states has two choices, and finally, they converge to one state in the bottom. 

However, we do not need to explore all of these states since they do not change the properties of interest, which only regards the physical environment such as the location of PLC machine parts. For example, suppose \(\mathit{PLC}_1\) moves one unit per time unit along the x-axis, starting from \((0,0)\), while \(\mathit{PLC}_2\) travels in parallel with $\mathit{PLC}_1$ from \((0,10)\). Regardless of the execution path taken, the system transitions from the state where \(\mathit{PLC}_1\) is at \((0,0)\) and \(\mathit{PLC}_2\) is at \((0,10)\) to the state where \(\mathit{PLC}_1\) reaches \((3,0)\) and \(\mathit{PLC}_2\) reaches \((3,10)\). Thus, instead of exploring 8 states, we can only explore one path consisting of 4 states.

Partial order reduction (POR) is a model checking optimization technique that reduces the number of explored states by identifying independent transitions that can occur in different orders without affecting the final system behavior. By selectively exploring only a subset of execution sequences, POR mitigates state explosion while preserving the correctness of verification results. We apply the ample-set based POR technique. By using our ample set approach, the state explosion only takes the left-most path in \Cref{fig:portransition}.  

A transition is specified by a rule label and substitution. When $s \xrightarrow{\alpha} s'$ in our semantics, $s'$ is the outcome of applying $l$ with $\sigma$ to $s$ for some rule label $l$ and a substitution $\sigma$. We denote this transition as $l(\sigma)$. When a rule label alone can determine a single enabled transition, we omit the substitution. Similarly, we only need a part of the substitution as long as it singles out the possible transition. For example, \texttt{rcvData(O} $\leftarrow$ \texttt{P1)} is a transition of \texttt{rcvData} where \texttt{P1} is the recipient. For simplicity, $\mathit{rule}\ \mathit{label}\texttt{(O} \leftarrow \texttt{P)}$ is abbreviated as $\mathit{rule}\ \mathit{label}(\texttt{P})$ for some object \texttt{P}. A timed transition of duration $\tau$ is represented as \texttt{tick}$(\tau)$.

We exploit the fact that physical behaviors for a scan cycle are determined by the actuation, which is dependent on the outcome of the previous cycle. We assume that actuation does not happen mid-cycle, which is a common practice in various PLC implementations. Thus, all the in-program transitions are independent of the external environment's time evolution \texttt{tick} and the starting of the new scan cycle \texttt{start} except for the communication functions. In addition, programs in different PLCs do not affect each other except for communication, since their transitions take place in strictly separate objects. Since there are independent groups of transitions, POR is suitable for reducing the state space.

Another major source of the state explosion is communication. Communication transitions can be mixed up with other transitions, such as assignments, start, tick, etc. However, communication happens independently of other types of transitions. Thus, we can allow the interleaving only among communication transitions. 

We also define internal transitions for individual PLCs (those explained in \Cref{fig:base-k}), which do not involve any inter-PLC interaction. Since these transitions and communication transitions are independent, we only need to consider the interleaving among the same groups. \Cref{def:transitions} are formal definitions of those groups of transitions.




\begin{definition}[Transition sets]
\label{def:transitions}
(1) Given a PLC object $P$, its internal transition set $P_{internal}$ contains transitions that modify \texttt{proc} of $P$. (2) The communication transitions set $\mathit{comm} = \{\texttt{sendData(P)}, \texttt{rcvData(P)}, \texttt{rcvFail(P)}, \texttt{rcvNo(P)}, \\ \texttt{isConnected(P)}, \texttt{connectRequest(P)} \mid \mbox{ for all PLC object } P\}$.
\end{definition}

Our ample set definition is as follows. The basic idea is that \texttt{start} is in the ample set when enabled. When \texttt{start} is not enabled, the ample set contains transitions of the object with the lowest index with at least one enabled transition. $\iota : \mathit{id} \rightarrow \mathbb{N}$ is a one-to-one numbering function that takes PLC identifiers and assigns distinct natural numbers to each PLC.

\begin{definition}[Ample set]
\label{def:ourample}
For a system $\mathcal{S}$ and a state $s$,
$\mathit{ample}$ is defined as follows:
(1) $\mathit{ample}(s) = \{\texttt{start}\}$, if $\mathit{start} \in \mathit{enabled}(s)$.
(2) $\mathit{ample}(s) =  (P_\mathit{internal} \cap \mathit{enabled}(s))$, if $\texttt{start} \notin \mathit{enabled}(s)$ and $\iota(P)$ is the minimal number such that $P_\mathit{internal} \cap \mathit{enabled}(s)$ is not empty.
(3) $\mathit{ample}(s) = \mathit{comm} \cap \mathit{enabled}(s)$ if $\texttt{start} \notin \mathit{enabled}(s)$ and $\mathit{comm} \cap \mathit{enabled}(s) \neq \emptyset\ \mbox{and } P_\mathit{internal} = \emptyset \mbox{ for all } P$.
(4) $\mathit{ample}(s) = \mathit{enabled}(s)\ \mbox{otherwise}$.
\end{definition}

Ultimately, we need to prove that the above definition of \texttt{ample} satisfies the four ample set conditions explained in \Cref{sec:prelim}. To achieve this goal, we first prove the independence between groups of transitions.

\begin{lemma}
\label{lem:independences}
The followings hold: (1) For any state $s$ where $\texttt{start} \in \mathit{enabled}(s)$, then \texttt{start} is independent of all transitions in $\mathit{enabled}(s)$.
(2) For any state $s$, \texttt{tick} is independent of all transitions in $\mathit{enabled}(s)$.
(3) For any two objects $P_1$ and $P_2$, for any $a_1$ and $a_2$ such that $a_1 \in \mathit{transition}_{P_1}$, $a_2 \in \mathit{transition}_{P_2}$, $a_1$ and $a_2$ are independent.
(4) Let $P$ be the object with the minimum numbering function in the system. Then, for any $a \in P_\mathit{internal}$ and $c \in \mathit{comm}$ are independent.
\begin{proof}
    (1) If \texttt{start} rule modifies a PLC object $P$, then any of $P$-specific transition are not enabled by definition of \texttt{start}.
    (2) \texttt{tick} rule updates the time and continuous attribute of PLC objects. The continuous attribute's behavior is only affected by the previous cycle of those PLC objects, so they are independent of transitions regarding PLC objects in the current cycle.
    (3) When $a_1$ and $a_2$ are non-communication transitions, let $P_1 \xrightarrow{a_1} P_1'$ and $P_2 \xrightarrow{a_2} P_2'$. Then, $P_1\ P_2\ \mathit{Conf} \xrightarrow{a_1} P_1'\ P_2\ \mathit{Conf} \xrightarrow{a_2} P_1'\ P_2'\ \mathit{Conf}$ and $P_1\ P_2\ \mathit{Conf} \xrightarrow{a_2} P_1\ P_2'\ \mathit{Conf} \xrightarrow{a_1} P_1'\ P_2'\ \mathit{Conf}$.
    (4) If \( c \) is not a communication transition related to \( P \), the lemma holds trivially. In the remaining case, the statement is vacuously true since transitions in \( P_\mathit{internal} \) and communication transitions involving \( P \) cannot be enabled simultaneously by construction of our semantics, where no two instructions are enabled at the same time. \qed
\end{proof}
\end{lemma}

Using the \Cref{lem:independences}, we have \Cref{thm:thm} stating that the ample set definition in \Cref{def:ourample} satisfies all the conditions required to be met when taking the ample set approach.

\begin{theorem}
\label{thm:thm}
\texttt{ample} defined in \Cref{def:ourample} satisfies the four conditions of partial order reduction.
\begin{proof}
    (1) Immediately follows from \Cref{def:ourample}. There is no case where $\mathit{enabled}(s)$ is an empty set, since \texttt{tick} is always enabled.
    (2) When $\mathit{ample}(s) = \{ \texttt{start} \}$, then by (1) of \Cref{lem:independences}, there is no dependent transition enabled. When  $\mathit{ample}(s) = (P_\mathit{internal} \cap \mathit{enabled}(s))$ for some PLC $P$, by \Cref{lem:independences}, all transitions that are dependent on any transition in $\mathit{ample}(s)$ are transitions in $P_\mathit{internal}$. 
    When  $\mathit{ample}(s) = (\mathit{comm} \cap \mathit{enabled}(s))$ or $\mathit{ample}(s) = \texttt{tick}$, there is no dependent transition with transitions in $\mathit{ample}(s)$ by \Cref{def:ourample} and (2) of \Cref{lem:independences}.
    (3) The property of interest is about continuous attributes, which are only affected by \texttt{tick} rule. Thus, all rules except \texttt{tick} rule are invisible. For any $s$, $s$ is not fully expanded iff $\texttt{tick} \in \mathit{enabled}(s)$, since \texttt{tick} is always enabled. Thus, when not fully expanded, transitions in $\mathit{ample}(s)$ are all invisible for any $s$.  
    (4) By construction of our semantics, the cycle in the transition system can occur only with an infinite loop. Since \texttt{while} is only in the ample set when fully expanded, the cycle in a transition system contains at least one fully expanded state. \qed
\end{proof}
\end{theorem}




\section{Formal Analysis using Rewriting modulo SMT}
\label{sec:separation}

\subsection{Symbolic Semantics}

Two key elements can be represented using SMT terms: the values stored in the \texttt{proc} attribute (internal) and the values mapped in \texttt{state} (external). The internal SMT constraints govern conditional statements and minimum message delay constraints, and the external SMT constraints reason the conformance of properties of the physical environment. 

Generally, the external constraints are heavier to solve because they handle the flow functions, whereas the internal constraints are confined to Boolean or linear constraints. In the semantics introduced so far, \texttt{tick} evolves time for both inside and outside the PLC programs. This introduces inefficiency of SMT solving, because not both types of constraints have to be checked every time \texttt{tick} applies.

We can introduce two separate clocks—one for the environment and another for the programs to address this problem. As discussed in the first part of \Cref{sec:reduction}, the physical behavior within a scan cycle is dictated by the results of the previous cycle, meaning that external behavior for the current cycle is already established at its onset. Consequently, the physical environment can progress continuously until the end of the current cycle. The internal clock must advance independently of the environmental clock to account for message delays. With this clock separation method, we can achieve higher efficiency while preserving the same result on analyses regarding the endpoints of scan cycles.  

To accomplish this, an additional timer is required to track the progression of time in the physical environment. The \texttt{PLCMachine} class has new attributes \texttt{envTimer} and \texttt{constraints}. \texttt{envTimer} contains the external time that can elapse without a change in physical dynamics and \texttt{constraints} collects the internal constraints for the enclosing PLC's programs.\footnote{Internal constraints can be gathered inside \texttt{proc} attribute, but we make a separate attribute for better presentation. The external constraints do not have to be explicitly declared for model checking. For example, we can encode the physical properties to be checked as Maude \texttt{search} command's conditions.}
\begin{maude}
class PLCMachine | state: Map{Id, Val}, flow: Flow, 
                  envTimer: Time, timer: Time, constraints: BooleanExpr .
\end{maude}

\begin{figure}[t]
    \centering
    \includegraphics[width=0.7\linewidth]{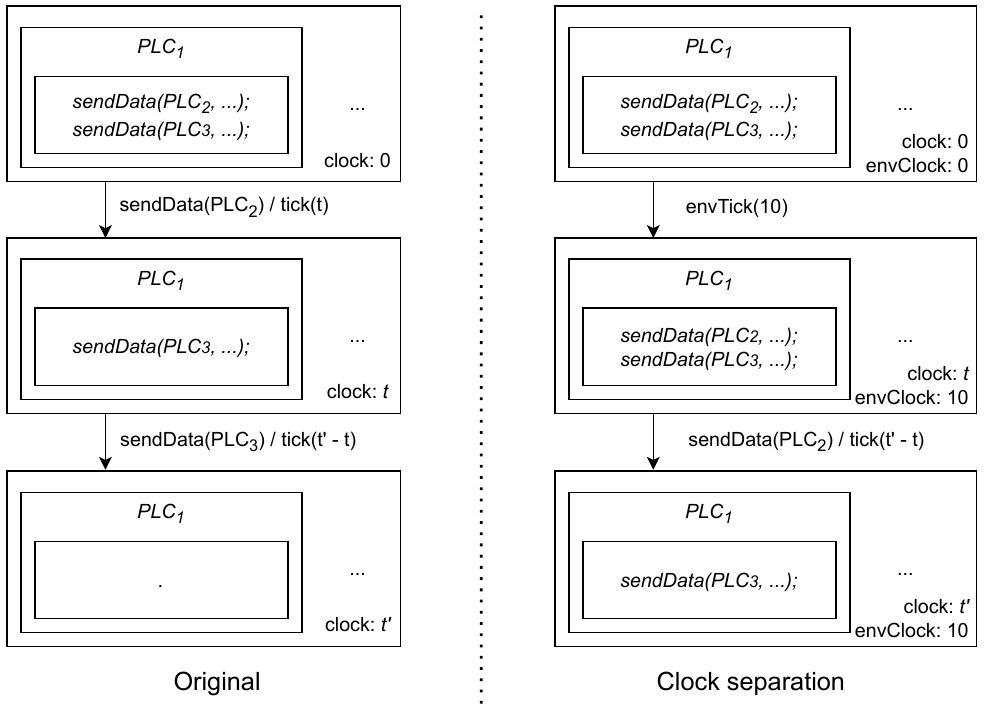}
    \caption{State transition before and after applying clock separation}
    \label{fig:clock-separation}
\end{figure}

\Cref{fig:clock-separation} presents two state transition diagrams. The left diagram illustrates the transitions without clock separation, while the right diagram depicts the transitions with clock separation. The system consists of $\mathit{PLC}_1$, $\mathit{PLC}_2$, and $\mathit{PLC}_3$ with scan cycle duration 10, and $\mathit{PLC}_1$ sends data to $\mathit{PLC}_2$ and $\mathit{PLC}_3$ ($\mathit{PLC}_2$ and $\mathit{PLC}_3$ are omitted in the figure for brevity). The system must take at least two transitions to distinguish whether two sent messages can be received. It is important to note that SMT solving costs regarding the environment is much higher than solving internal constraints. On the left, where the clock is not separated, we must solve the internal and external constraints for all the tick rules. However, on the right side, the environmental clock can immediately jump to 10, solving the SMT constraints for the physical environment just once, and internal time jumps (\texttt{tick}) only regard the internal constraints: message delay constraint in this case.

\texttt{tick}, \texttt{timeEffect}, and \texttt{mte}'s definitions for symbolic execution are shown below. They are now only for internal time elapses. \texttt{freshVarGen} generates a fresh SMT variable to denote the duration of newly elapsed time. \texttt{mte} returns Boolean expression that the newly elapsed time is less than the maximal time elapse of the current system. \texttt{addConst} adds the constraints to all PLC objects.
\begin{maude}
var PLC : Object .   var S : Stmt . var O : Oid .
var CONF : Configuration .          vars T TIMER C : Time .
vars CONST BE : BooleanExpr .       vars V1 V2 : Val .
var ENV : Map{Id, Loc} .            var STORE : Map{Loc, Val} .
vars STATE R1 R2 : Map{Id, Val} .   vars ATTRS1 ATTRS2 : AttributeSet .

 rl [tick] : { CONF } => { addConst(timeEffect(CONF, T), mte(CONF)) } 
 if T := freshVarGen(CONF) .

 eq mte(PLC CONF, T) = mte(PLC, T) AND mte(CONF, T) .
 eq mte(< O : PLCMachine | timer : TIMER >, T) = T <= TIMER .
 eq mte(CONF, T) = true [owise] .
 eq addConst(PLC CONF, COSNT) 
  = addConst(PLC, CONST) addConst(CONF, CONST) .
 eq addConst(CONF, CONST) = CONF [owise] .
 eq addConst(< O:Oid : PLCMachine | constraints : CONST >, CONST')
  = < O:Oid : PLCMachine | constraints : CONST and CONST' > .
\end{maude}

The \texttt{constraints} attribute also collects the constraints from executing conditional statements.
\begin{maude}
crl [if-true] : 
   < O:Oid : PLCMachine | proc : k(if BE then S else S' ~> ...) ...,
                          constraints : CONST >
=> < O:Oid : PLCMachine | proc : k(S ~> ...) ...,
                          constraints : CONST and BE >
if smtCheck(CONST and BE) .

crl [if-false] : 
   < O:Oid : PLCMachine | proc : k(if BE then S else S' ~> ...) ...,
                          constraints : CONST >
=> < O:Oid : PLCMachine | proc : k(S' ~> ...) ...,
                          constraints : CONST and not BE >
if smtCheck(CONST and not BE) .
\end{maude}

The start rule\footnote{it should be a rewrite rule instead of an equation to give the system nondeterministic choice whether to apply \texttt{start} or not when enabled.} now checks that the current timer value \texttt{C} \emph{can be} zero, accounting for the accumulated constraints \texttt{CONST}.
\begin{maude}
crl [start] :
 start(< O : PLCMachine | timer : C, 
                          proc : [cycleTime(T) env(ENV) store(STORE) KC], 
                          state : STATE, constraints(CONST) >)
     = < O : PLCMachine | timer : C, proc : cycleTime(T) env(ENV)
                          store(sense(ENV, STORE, STATE)) KC,
                          state : actuate(ENV, STORE, STATE) >
 if smtCheck(C == 0 and CONST) .
\end{maude}

Similarly, \texttt{rcvData} is modified to check whether the message can be accepted.
\begin{maude}
crl [rcvData] :
    < O : PLC | proc : k(rcvData(O', SFBID, RFBID) ~> K) KC >
    < link(O, O') : Channel | validity : true, 
            buffer : (BUFFER m(O', O, SFBID, RFBID, DATA, DTIMER)) >
 => < O : PLC | proc : k(DATA ~> K) KC >
    < link(O, O') : Channel | validity : true, buffer : BUFFER >
 if smtCheck(DTIMER <= 0) .
\end{maude}

The physical part of \texttt{tick} rule introduced in \Cref{sec:external} is separated to \texttt{envTick} rule, when \texttt{minEnv} returns the smallest time duration to reach any end of the PLCs' scan cycles. \texttt{smtCheck} checks if the given SMT formula is satisfiable (supported in Maude). \texttt{checkProperty} is a function that generates an SMT constraint implying the negation of desired property of the system.  
\begin{maude}
crl [envTick] : {CONF} => {envTimeEffect(CONF, minEnv(CONF))}
 if minEnv(CONF) > 0 .

 eq minEnv(< O : PLCMachine | envTimer : TIMER, flow : FLOW, 
              state : STATE > CONF) = min(TIMER, minEnv(CONF)) .
 eq minEnv(none) = infinity .
\end{maude}
Accordingly, \texttt{envTimeEffect} is defined. It updates \texttt{envTimer} instead of \texttt{timer}.
\begin{maude}
eq envTimeEffect(
   < O : PLCMachine | envTimer : TIMER, flow : FLOW, state: STATE >, T)
 = < O : PLCMachine | envTimer : monus(TIMER, T), 
                      state : eval(FLOW, STATE, T) > .
\end{maude}

\subsection{Formal Analysis}
\label{sec:formalAnalysis}

We aim to run a reachability analysis on the chemical plant with two tanks explained in \Cref{sec:motiv}. Consider the following initial state. The \texttt{clock} attribute is newly introduced to represent the total elapsed system time, allowing for the specification of the analysis scope in terms of system time.
\begin{maude}
var init : Configuration .
eq init =
< "plc1" : PLCMachine | proc : app1, timer : 0, envTimer : 0, clock : 0
                        senState : water_level |-> 20,
                        actState : pump_switch |-> 0,
                flow : water_level(t) = water_level - pump_switch *t >
< "plc2" : PLCMachine | proc : app2, timer : 0, envTimer : 0, clock : 0
                        senList : water_level(t) |-> 20,
                        actList : pump_switch |-> 0,
                flow : water_level(t) = water_level - pump_switch * t >
< conn("plc1", "plc2") : Conn | 
        validity : false, buffer : emptyBuffer > .
\end{maude}

We want to check if the water level always stays between 2 and 35 within the time bound 20 ms. The time bound is enforced by adding an equation that converts the whole system into an operator that cannot be rewritten again. 
\begin{maude}
 op boundReached : GlobalSystem [ctor] .
ceq { < O:Oid : PLCMachine |  ATTRS, clock : T > Conf } = boundReached . 
 if T > 20 .
\end{maude}

The following \texttt{search} command searches for a reachable state that goes outside the specified range of water level. The result suggests that both tanks in the system maintain the specified water levels up to 20 time units.
\begin{maude}
search [1] init =>*
{< "plc1" : PLCMachine | state : (water_level |-> V1, R1), ATTRS1 >
 < "plc2" : PLCMachine | state : (water_level |-> V2, R2) ATTRS2 > CONF }
such that smtCheck(V1 < 2 or V1 > 35 or V2 < 2 or V2 > 35) .
No solution.
\end{maude}

Now, we want to check that the two tanks have exactly the same water levels from the state with unbalanced water levels. Consider the following initial state.
\begin{maude}
eq init =
< "plc1" : PLCMachine | proc : app1, timer : 0, envTimer : 0, clock : 0
                        state : water_level |-> 5, pump_switch |-> 0,
                flow : water_level(t) = water_level - pump_switch * t >
< "plc2" : PLCMachine | proc : app2, timer : 0, envTimer : 0, clock : 0
                        state : water_level(t) |-> 45, pump_switch |-> 0,
                flow : water_level(t) = water_level - pump_switch * t >
< link("plc1", "plc2") : Channel |
        validity : false, buffer : emptyBuffer > .
\end{maude}

The following search command looks for the reachable state where the two water levels are the same. This search command finds a reachable state and presents it to the user. \texttt{ATTRS1} and \texttt{ATTRS2} are omitted for brevity. 
\begin{maude}
search [1] init =>* 
{< "plc1" : PLCMachine | senState : water_level |-> V1, ATTRS1 > 
 < "plc2" : PLCMachine | senState : water_level |-> V2, ATTRS2 >} 
such that smtCheck(V1 === V2) .
Solution 1 (state 828)
states: 829  rewrites: 160521 in 2220ms cpu (2220ms real)
(72306 rewrites/second)
ATTRS1 --> ...   V1 --> 25
ATTRS2 --> ...   V2 --> 25
CONF --> < link("plc1", "plc2") : Channel | validity : true,
             buffer : emptyBuffer >
\end{maude}


\section{Experimental Evaluation}
\label{sec:expr}

To assess the effectiveness of our approach, we implemented our semantics and state space reduction techniques in Maude~\cite{maude-manual}, a high-performance rewriting engine. Nondeterministic time evolution is handled using symbolic execution methods, such as those described in~\cite{plcst2022lee}.
We developed a total of eight benchmark models using both our semantics and the SpaceEx model. In our framework, modeling requires minimal effort, as the PLC ST code itself serves as the model; only the physical environment and \texttt{Conn} settings need to be configured. In contrast, the SpaceEx model requires manually constructing all models from scratch.

The first research question examines whether our approach is more efficient than the previous hybrid automata-based approach using SpaceEx. To evaluate this, we compare the time required for full-state exploration in Maude and SpaceEx.
The second research question regards the effectiveness of our state space reduction technique. We compare the space and time spent in full-state exploration before and after applying the state space reduction techniques. \Cref{sec:benchmarks} describes the benchmark models, \Cref{sec:externalExpr} investigates the first research question, and \Cref{sec:internal} examines the second research question. All experiments were conducted on Intel Xeon 2.8 GHz with 256 GB memory. Timeout is set to 10 minutes in all settings.

\subsection{Benchmark Models}
\label{sec:benchmarks}

Previous benchmark sets for PLC programs are often limited in scope, focusing primarily on isolated control logic without modeling inter-PLC communication or physical dynamics. Many benchmarks consist of single-task programs. As a result, they fail to capture essential aspects of PLC systems, such as concurrent execution, communication, and interaction with continuous physical processes.

We obtain models with physical dynamics, programming logic, and inter-PLC communication by adapting hybrid automata benchmarks. They often include both discrete control logic and continuous physical dynamics by themselves, and it is possible to encode simplified message-passing logic. 

Besides hybrid automata, we construct the secure water treatment (SWaT) model, which is a standard industrial control system model. They have communication, physical environment, and programming logic altogether. Our work is the first to construct comprehensive SWaT models, which closely model a real-world example, that can be utilized in formal analysis to the best of our knowledge.

We have eight benchmark models, including a chemical plant with two pumps \cite{nellen2015,nellen2016}, a railed vehicle, and networked thermostats \cite{bae2016smt}, each implemented with and without explicit communication. On top of that, we have two models for two parts of SWaT process \cite{minicps}.

In the models without explicit communication, multiple PLCs are consolidated into a single PLC that runs multiple programs and encapsulates all physical behaviors. In contrast, the models with explicit communication maintain multiple PLC objects that exchange data and operate accordingly. \emph{Two tanks with two pumps} models are already covered in \Cref{sec:motiv}, \Cref{sec:comm-example}, and \Cref{sec:formalAnalysis}.

\paragraph{Railed Vehicles.}
\begin{figure}
    \centering
    \includegraphics[width=0.5\linewidth]{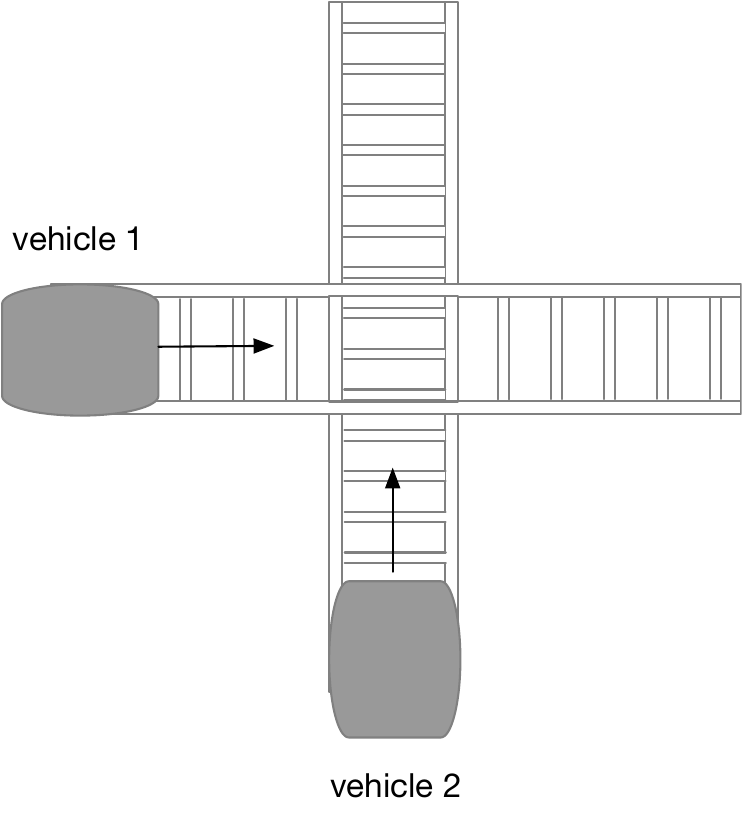}
    \caption{Railed vehicle model diagram}
    \label{fig:railed-vehicle}
\end{figure}
The railed vehicle model depicted in \Cref{fig:railed-vehicle} represents a system in which two autonomous vehicles move along a fixed rail track while coordinating their movements to avoid collisions. The track contains a critical intersection at (10,10) where both vehicles' paths overlap, making communication between them essential for safe operation.

The first vehicle starts at position (0,10) and moves along the horizontal axis toward (20,10), while the second vehicle starts at (10,0) and moves along the vertical axis toward (10,20). Both vehicles receive Boolean user inputs that dictate their movement: a True input instructs the vehicle to proceed, while a False input causes it to stop. Since the intersection at (10,10) poses a risk of collision, the vehicles exchange their user inputs before moving.
If both vehicles receive a True input simultaneously, vehicle 1 takes precedence and signals vehicle 2 to halt, preventing a potential collision at the intersection.

This model captures the core principles of distributed coordination and collision avoidance in cyber-physical systems. It demonstrates how PLC-based automation can handle real-time decision-making through communication protocols, ensuring safe and efficient operations in environments where multiple entities share limited resources.

\paragraph{Networked Thermostat.}
The networked thermostat model represents a temperature control system for two isolated rooms, each equipped with a heater that regulates its temperature. The model captures both the natural heat loss that occurs over time and the heater activation logic used to maintain a desired temperature range while optimizing energy efficiency.

Regardless of whether the heater is active, each room experiences heat loss every time unit, causing a gradual temperature decrease. When a room’s heater is turned on, its temperature increases over time, counteracting the cooling effect. The system monitors the sum of the temperatures of both rooms to determine the appropriate heating strategy.

The heater activation follows a two-step control policy. When the combined temperature of both rooms falls below a predefined threshold, both heaters activate simultaneously. However, when the total temperature is above the threshold, only the heater in the cooler room turns on. This ensures that the system maintains a comfortable temperature while minimizing energy consumption by avoiding unnecessary simultaneous heater operation.

This model reflects energy-efficient temperature regulation strategies commonly used in building automation and smart thermostat systems. By dynamically adjusting heater usage based on real-time temperature conditions, the system balances comfort and efficiency, effectively representing intelligent heating control in cyber-physical systems.

\paragraph{SWaT.}
The SWaT (Secure Water Treatment) benchmark is a widely recognized testbed simulating a real-world water treatment facility composed of six sequential stages. For the purpose of our analysis, we focus on control-oriented logic and exclude stages that primarily involve chemistry-related processes, such as pH adjustment and chemical dosing. Our implementation is composed of two models, based on the simplified SWaT model found in of the work \cite{minicps}. The first model governs the operation of the water inlet valve and the backwash filter. The second model incorporates communication between PLCs to coordinate operations based on the system's state.

The first model is similar to our motivating example in \Cref{sec:motiv}. It includes conditional statements that open or close the inlet pump if the water level is over or under predefined thresholds. It keeps the current valve status if the water level is within the desired range. There is a constant outflow of water from the tank. The second model decides the outflow of the model based on the communication between PLCs. There are three networked PLC machines: the first one includes the main control of the pumps and reflects the communication output into actuators, while the other two provides the control inputs.

\subsection{Time Comparison with Hybrid Automata-based Approach}
\label{sec:externalExpr}

We compare the full-state exploration time of our rewrite-based approach, which incorporates state space reduction techniques, with that of the SpaceEx tool. To enable a meaningful comparison, we manually constructed SpaceEx models that approximate the behavior of our approach in a simplified manner: mode transitions encode blocked code segments rather than one-step rewrites. On top of that, the complex behavior of communication function blocks such as connection checking and flag settings are completely abstracted out.

While our framework currently supports polynomial dynamics of arbitrary degree, for this experiment, we restrict the dynamics to linear functions to align with the capabilities of the SpaceEx tool under the PHAVer scenario.

\begin{table}[t]
    \centering
    \caption{Time Comparison between our Approach and SpaceEx}
    \includegraphics[width=1.0\linewidth]{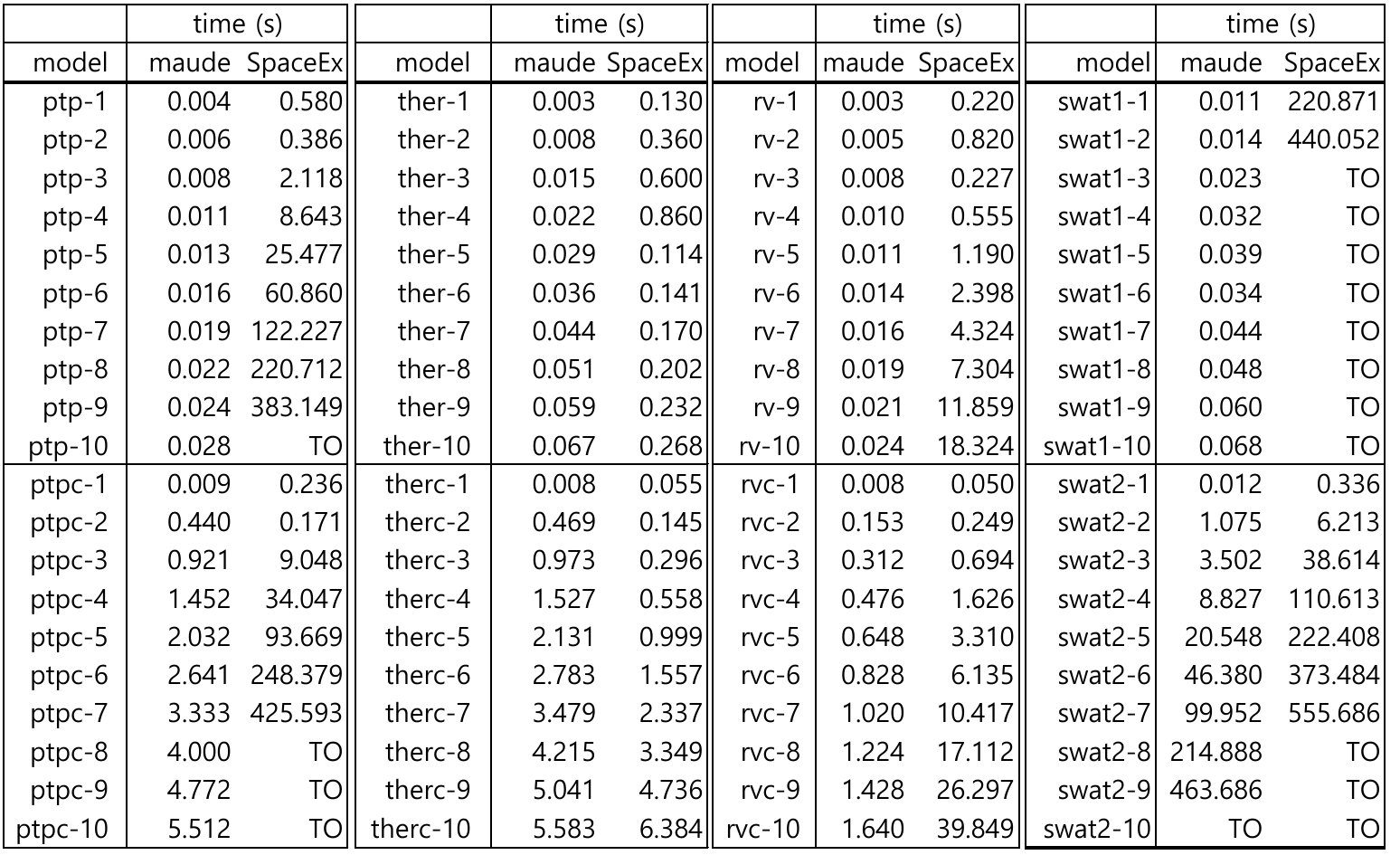}
    \label{table:external}
\end{table}

\Cref{table:external} presents the execution time comparison between our approach and SpaceEx, measured in milliseconds. Each model is identified using the format \emph{model name}-\emph{time bound}. The model names 'ptp', 'ther', and 'rv' correspond to a chemical plant with two pumps, a thermostat, and a railed vehicle without communication, respectively. When 'c' is appended to the model name, it indicates that the model includes explicit communication. The time bound is expressed as a multiple of the scan cycle duration. For instance, 'ptpc-3' represents a chemical plant with two pumps, explicit communication, and a time bound of three scan cycles. Also note that analyses with the model 'ptpc-*' is covered in \Cref{sec:formalAnalysis}. 'swat1' and 'swat2' correspond to the first and second halves of the SWaT process. 

The execution time comparison between our approach (denoted 'maude' in the table) and the SpaceEx tool shows significant differences in efficiency, particularly as the system complexity increases. Across all benchmark models, our approach consistently exhibits lower execution times.

For models without explicit communication, the execution time for our approach remains relatively low even as the time bound increases. In contrast, the SpaceEx execution time grows exponentially, particularly for ptp-* models, where the execution is timed out at ptp-10, compared to 0.028 s in our approach. This trend is evident across all models, demonstrating how our method effectively controls state space growth while maintaining accuracy.

For models with explicit communication, a natural increase in execution time is observed due to the additional complexity of message passing. However, our approach still performs significantly better than SpaceEx. 'ptpc' shows the greatest time gap between our approach and SpaceEx. SpaceEx model is timed out from 'ptpc-8', where it only takes 3.063 seconds with our approach. The thermostat model (therc-*) exhibits a similar pattern. Although SpaceEx outperform our approach in some model bound settings, our approach eventually outperforms SpaceEx again as the bound gets larger. This result suggests that our rewrite-based approach is more promising for large-scale industrial control systems than SpaceEx.

In \cite{nellen2016}, their approach for the chemical plant with two pumps without communication takes 69.0 seconds to complete the analysis on their machine.

\subsection{Effectiveness of State Space Reduction}
\label{sec:internal}

\begin{figure}[t]
    \centering
    \includegraphics[width=1.0\linewidth]{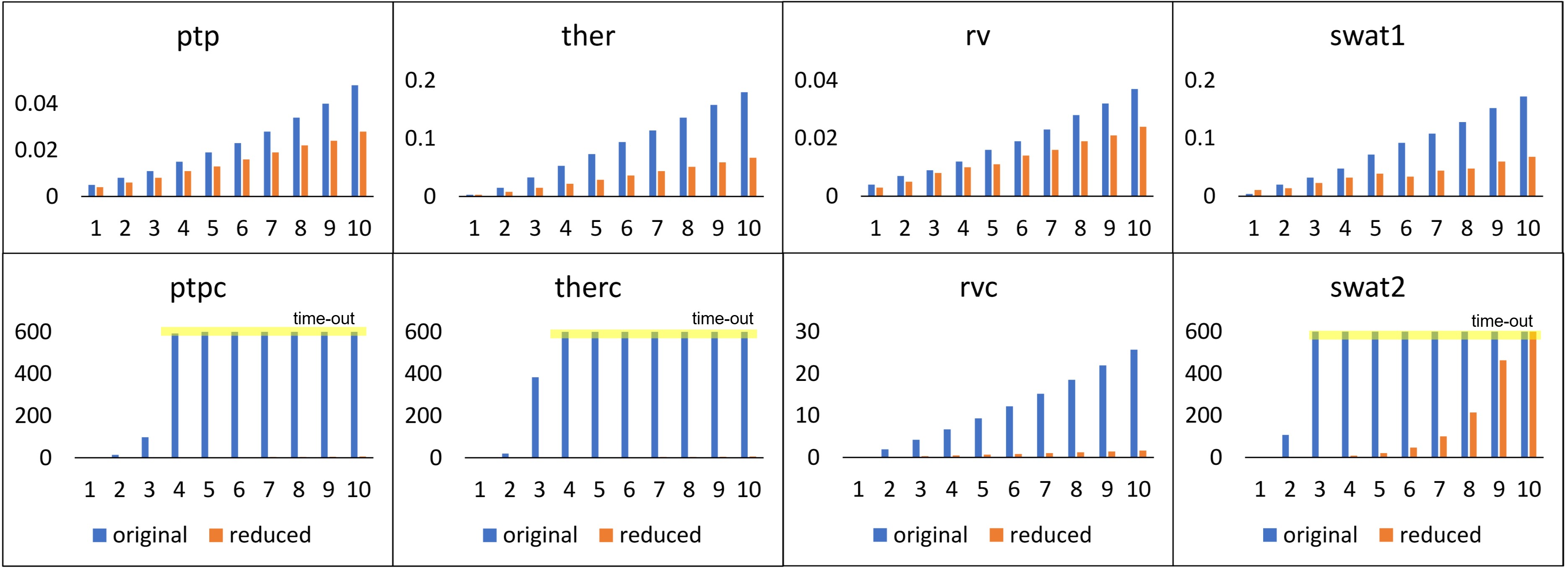}
    \caption{Analysis Time Comparison before and after State Space Reduction}
    \label{fig:graphs}
\end{figure}

To demonstrate the effectiveness of our state space reduction techniques, we compare the full-state space exploration time before and after their application. \Cref{fig:graphs} shows 8 graphs that compare the original and reduced semantics analysis time. For each graph, the x-axis is the increasing time bound of the analysis setting from 1 cycle time to 10 cycle times. The y-axis shows the time taken for the full state exploration in seconds. The timed-out data are shown as 600 seconds. \Cref{table:internal} presents a comparison of space and time usage in our approach, with and without the state space reduction technique. Execution time is measured in milliseconds, while space is quantified by the number of symbolic states explored. 'TO' means timeout and '-'s are placeholder for the space size for the timed-out model settings. The model identifiers remain consistent with those used in \Cref{table:external}. In the top row, \texttt{original} and \texttt{reduced} indicate the values before and after applying state space reduction, respectively.

\begin{table}[]
    \centering
    \caption{Space and Time Comparison Before and After State Space Reduction}
    \includegraphics[width=1.0\linewidth]{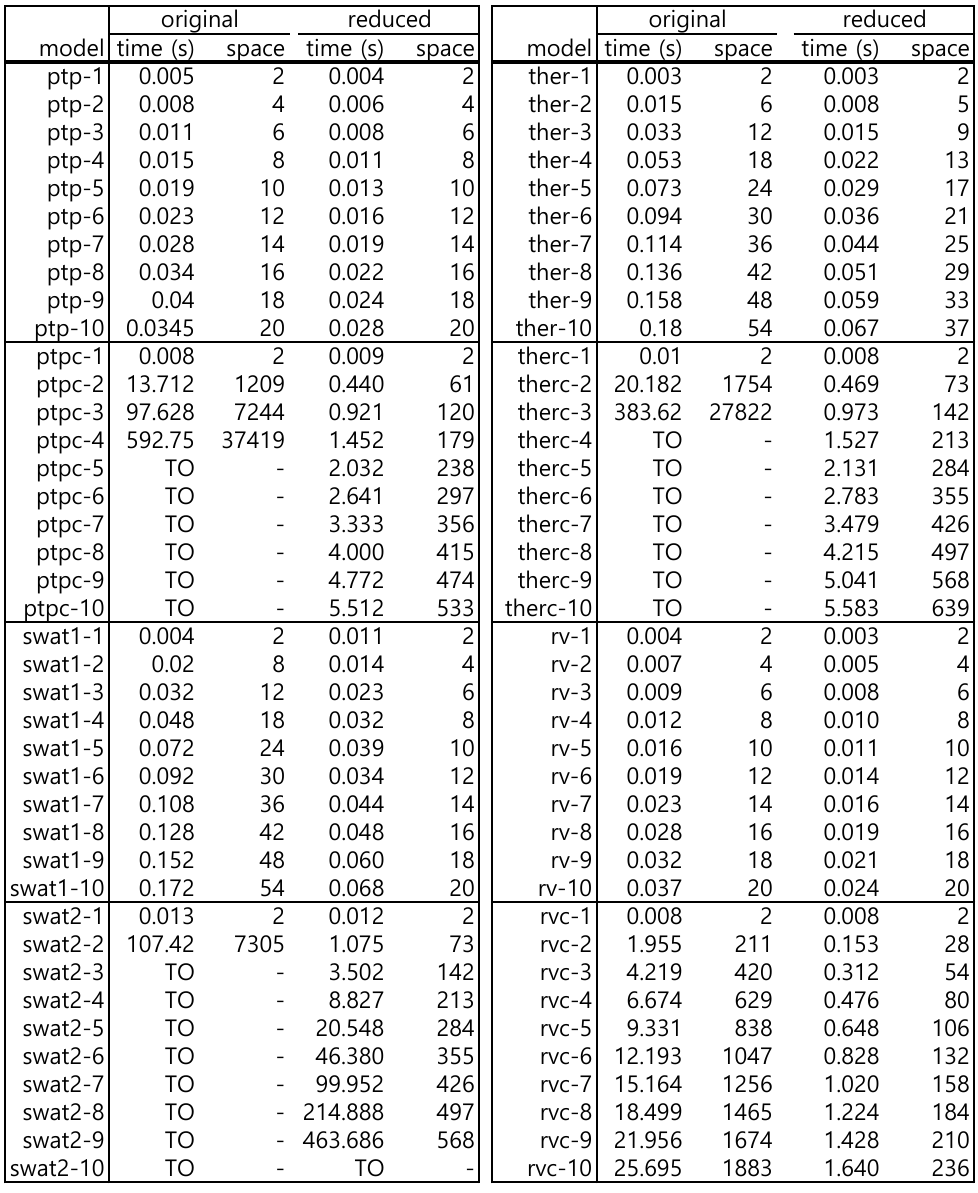}
    \label{table:internal}
\end{table}

The results in \Cref{fig:graphs} highlight the impact of state space reduction techniques. Across all benchmark models, applying state space reduction consistently reduces execution time, with significant improvements observed in models that involve explicit communication.

For models without communication, execution time is mildly lowered, visually expressed as the original (blue) bars are longer than the reduced (orange) bars.
For models with explicit communication, the benefits of state space reduction are more pronounced. The orange bars are not even visible in some cases. 
Across all communication-based models, the original approach results in rapid state space growth, leading to timeouts (TO) in larger instances, whereas the reduced version remains computationally feasible. The blue bars that reach 600 in 'ptpc', 'therc' and 'swat2' are timed out data before state space reduction.

These results demonstrate that state space reduction is crucial for handling complex PLC-based systems with communication, where nondeterminism plays a significant role in increasing verification complexity.

\section{Concluding Remarks}
\label{sec:concl}

In this work, we have presented a formal semantics for industrial control systems modeled using programmable logic controllers (PLCs). 
Our framework integrates PLC program execution, interactions with the physical environment, and networked communication, enabling 
a unified analysis of system behavior. By incorporating rewriting logic, we provide a modular and expressive formalization that faithfully 
represents real-world industrial automation scenarios.

To address the state explosion problem inherent in formal verification, we introduced a state space reduction technique based on partial order reduction,
 significantly improving analysis efficiency. Our approach is implemented in Maude and evaluated using benchmark models, demonstrating its scalability 
 and effectiveness. Compared to SpaceEx, our method requires less modeling effort while offering a more precise and scalable verification framework.

Future work includes extending our semantics to support more complex multitasking and real-time scheduling features, as well as further optimizing 
state space reduction techniques. Additionally, we aim to integrate our approach with existing PLC verification tools to provide a more comprehensive and practical formal analysis framework for industrial control systems.

\bibliographystyle{splncs04}
\bibliography{kbib}
\clearpage
\appendix
\end{document}